\documentclass[preprint,12pt,3p,review,sort&compress]{elsarticle}

\usepackage{microtype}
\usepackage{epstopdf}
\usepackage{amsmath}
\usepackage{amssymb}
\usepackage{array}
\usepackage{mdwmath}
\usepackage{mdwtab}
\usepackage{eqparbox}
\usepackage{multirow}
\usepackage[tight,footnotesize]{subfigure}
\usepackage{stfloats}
\usepackage[bookmarks=true,hidelinks]{hyperref}
\usepackage{bookmark}
\usepackage{setspace}

\begin{document}

\begin{frontmatter}



\title{Deep Fidelity in DNN Watermarking: A Study of\\Backdoor Watermarking for Classification Models}

\author[label1]{Guang Hua}\ead{ghua@ieee.org, guang.hua@singaporetech.edu.sg}
\author[label2]{Andrew Beng Jin Teoh\corref{correspond}}\ead{bjteoh@yonsei.ac.kr}
 
\address[label1]{Infocomm Technology (ICT) Cluster, Singapore Institute of Technology (SIT), Singapore 138683, Singapore}
\address[label2]{School of Electrical and Electronic Engineering, Yonsei University, Seoul 120749, South Korea}

\cortext[correspond]{Corresponding author.}

\begin{abstract}
Backdoor watermarking is a promising paradigm to protect the copyright of deep neural network (DNN) models. In the existing works on this subject, researchers have intensively focused on watermarking robustness, while the concept of fidelity, which is concerned with the preservation of the model's original functionality, has received less attention. In this paper, focusing on deep image classification models, we show that the existing shared notion of the sole measurement of learning accuracy is inadequate to characterize backdoor fidelity. Meanwhile, we show that the analogous concept of embedding distortion in multimedia watermarking, interpreted as the total weight loss (TWL) in DNN backdoor watermarking, is also problematic for fidelity measurement. To address this challenge, we propose the concept of \emph{deep fidelity}, which states that the backdoor watermarked DNN model should preserve both the feature representation and decision boundary of the unwatermarked host model. To achieve deep fidelity, we propose two loss functions termed penultimate feature loss (PFL) and softmax probability-distribution loss (SPL) to preserve feature representation, while the decision boundary is preserved by the proposed fix last layer (FixLL) treatment, inspired by the recent discovery that deep learning with a fixed classifier causes no loss of learning accuracy. With the above designs, both embedding from scratch and fine-tuning strategies are implemented to evaluate the deep fidelity of backdoor embedding, whose advantages over the existing methods are verified via experiments using ResNet18 for MNIST and CIFAR-10 classifications, and wide residual network (i.e., WRN28\_10) for CIFAR-100 task. PyTorch codes are available at: \textbf{\url{https://github.com/ghua-ac/dnn_watermark}}.
\end{abstract}


\end{frontmatter}

\section{Introduction}
\label{sec_intro}
Deep learning technology has revolutionized many fields of modern society, from computer vision to natural language processing, from medicine to education, from individual life to smart nations, and everything in between \cite{Deep_Learning_Nature}. As the outcome of sophisticated research and development works enabled by human intelligence, high-quality big data, and cutting-edge computing power, among which none is free, deep neural network (DNN) models are a new form of digital intellectual property that calls for effective protection from copyright infringement. This leads to the emerging research topic of DNN watermarking \cite{Survey_2021_NN_Watermark,Funeral_2021}. 

Analogous to multimedia watermarking \cite{PR_2022_Watermarking} in which watermarks are embedded in terms of modifying host multimedia content, DNN models can be watermarked by modifying host model parameters, e.g., adding statistical bias \cite{Inside_2017_Uchida}, modifying layer outputs \cite{Inside_2018_ICMR}, binarizing transformed random weights \cite{Inside_2020_Compensate}, quantizing fully-connected layers \cite{Inside_2020_FC}, shifting histogram of low-entropy weights \cite{Inside_Reversible_2020_ACMMM}, etc. These methods are also known as \emph{internal embedding}. Alternatively, if considered as data processing systems, DNN models can be watermarked via \emph{backdoor embedding}, which establishes special model behaviors when queried by a set of so-called trigger samples  \cite{Backdoor_2018_AsiaCCS,Backdoor_2018_Signature,Backdoor_2018_turning_weakness,Backdoor_2019_Encoder,Backdoor_2019_Exponential,Backdoor_2020_Adversarial,Backdoor_2020_Hash,BD_new_label_2020,Backdoor_2020_New_Label,Backdoor_fragile_2021,Backdoor_2021_Usenix_Entangled,Backdoor_2022_with_Steg,Backdoor_2023_Holo} (details in Section 2). The two types of embedding methods can also be jointly implemented within a single framework for enhanced protection capability, \cite{Inside_2019_Passport}. In either form, the embedded watermarks can be extracted in piracy-suspicious or illegally distributed models and presented to the authorities for authentication, ownership verification, or traceback purposes. The existing DNN watermarking methods can be applied to both discriminative and generative DNN models \cite{Generative_2020_Image_Process_CSVT}, while recently the scope has been further extended to image captioning models \cite{Backdoor_PR_2022_Image_Captioning}, graph neural networks \cite{Backdoor_GNN_2021}, and federated learning \cite{Backdoor_Federated_2021}.

While extensive research attention has been drawn to the improvement of DNN watermark robustness against the existing attacks and other performance aspects, the fidelity issue, which characterizes how the watermarked model performs differently from the host model on the original task, has been treated in a relatively simplistic way. We note that in both the existing internal embedding and backdoor-based methods, e.g., \cite{Inside_2017_Uchida,Inside_2018_ICMR,Inside_2020_Compensate,Inside_2020_FC,Inside_2020_Interspeech_SS,Inside_Reversible_2020_ACMMM,Backdoor_2018_AsiaCCS,Backdoor_2018_Signature,Backdoor_2018_turning_weakness,Backdoor_2019_Encoder,Backdoor_2019_Exponential,Backdoor_2020_Adversarial,Backdoor_2020_Hash,BD_new_label_2020,Backdoor_2020_New_Label,Backdoor_fragile_2021,Backdoor_2021_Usenix_Entangled,Backdoor_2022_with_Steg,Backdoor_2023_Holo,Inside_2019_Passport,Inside_and_backdoor_2019_Deepsigns}, fidelity has been measured by the classification accuracy of the watermarked model on the learning set, wherein the accuracy drop with respect to the host model due to watermark embedding is supposed to be minimized. The underlying rationale lies in the special structure of DNN models in contrast to multimedia content, and it has been argued and widely accepted that the alteration of model parameters due to watermark embedding is less important as long as the classification accuracy is preserved \cite{Survey_2021_NN_Watermark}. However, this notion may lead to a few risks.
\begin{enumerate}
	\item It neglects internal mechanisms that can reflect functional characteristics in the forward path of the model. Because of the overparameterization (redundancy) and the randomness during training, the same model architecture with significantly different parameters can still yield the same learning accuracy. Therefore, merely using the accuracy is somewhat misleading and insufficient to measure watermark fidelity.
	\item Even a watermarked model with an unaltered learning accuracy (perfect fidelity under the current notion) can yield significantly different results when deployed in real-world environments. In other words, the black-box nature of DNN models will lead to non-explainable behaviors before and after watermark embedding if tested by unseen data, especially adversarial examples \cite{Survey_2020_Adversarial_Examples_Field} and open-set samples \cite{Survey_2020_Open_Set_Recognition}, even if the parameters are just slightly modified through watermark embedding.
	\item The analogy between multimedia and DNN watermark fidelity is still ambiguous and worth of further investigation for a rigorous definition of the latter.
\end{enumerate}

Therefore, it is necessary to look for more descriptive means in addition to the existing learning accuracy to characterize the behavior of a DNN model and thus more appropriately measure the fidelity. We note that in DNN models for image classification tasks, representation and classification are jointly learned through training. That is to say, the feature extraction layers, e.g., convolution-based modules, perform representation learning, which are typically followed by a multi-layer perception (MLP) with fully connected layers for classification. In the state-of-the-art architectures \cite{NIN,GoogLeNet,ResNet}, the MLP classifier is replaced by a single-layer linear classifier without performance degradation, thanks to the powerful feature learning capability of DNNs. In either case, we can view the model as the concatenation of two functional components, i.e., a feature extractor and a subsequent single-layer linear classifier. This notion has been widely incorporated in general deep learning research, e.g., fixed classifier \cite{Fix_Pro_2018_ICLR}, polytope networks \cite{Fix_Pro_2021_TNNLS}, etc., as well as other vision tasks, e.g., center loss \cite{Face1_ECCV2016}, CosFace \cite{Face2_CVPR2018}, and ArcFace \cite{Face3_CVPR2019} for face recognition.

From this perspective, we therefore propose the concept of \emph{\textbf{deep fidelity}}, which states that the watermarked DNN model, upon successful watermark embedding, should preserve both the feature representation and classification boundary of the host model. In doing so, model functionality can be concretized from the only learning accuracy to how the model performs feature extraction and makes decisions. Note that well-preserved feature extraction and decision boundary are the sufficient condition for well-preserved learning accuracy, but not vice versa, which reveals the advantage of the deep fidelity concept. Based on the notion of deep fidelity, to preserve feature representation, we propose two loss functions termed as penultimate feature loss (PFL) and softmax probability-distribution loss (SPL), respectively, to regularize the watermark embedding process. Meanwhile, to preserve decision boundary, inspired by the recent discovery that fixing the last layer classifier of a DNN model causes no loss of learning accuracy \cite{Fix_Pro_2018_ICLR, Fix_Pro_2021_TNNLS}, we propose the fix last layer (FixLL) treatment during watermark embedding. The above designs are then consolidated with both training from scratch and fine-tuning based strategies to realize backdoor embedding. While the notion of deep fidelity can be applied to both internal and backdoor watermarking of different types of DNN models, we focus on backdoor watermarking for deep image classification models in this paper. Our contributions are summarized as follows.

\begin{enumerate}
	\item We reveal the limitations of the existing shared learning-accuracy-based metric for DNN watermark fidelity and propose the novel concept of \emph{deep fidelity} as a solution.
	\item Focusing on deep discriminative models for image classification tasks, we further propose a series of techniques (FixLL, PFL, SPL, etc.) to realize deep fidelity.
	\item We have carried out extensive experiments in comparison with state-of-the-art methods, rigorously verifying the advantages and practical merits of our proposal.
\end{enumerate}

The paper is organized as follows. In Section \ref{sec_related}, we comprehensively review the existing backdoor watermarking methods, emphasizing the embedding strategies therein. In Section \ref{sec_3}, we analyze the limitations of the existing shared definition of backdoor fidelity from the analogy of multimedia watermarking fidelity and also from the perspective of model behavior. Then, we propose the concept of deep fidelity in Section \ref{sec_4}, based on which several schemes for deep fidelity backdoor embedding are designed. The performance of the proposed deep fidelity embedding schemes is evaluated in Section \ref{sec_5} through extensive analysis and experimental results. Finally, Section \ref{sec_6} concludes the paper.

\section{Related Work}
\label{sec_related}
In this section, we review related works on DNN backdoor watermark embedding \cite{Backdoor_2018_AsiaCCS,Backdoor_2018_Signature,Backdoor_2018_turning_weakness,Backdoor_2019_Encoder,Backdoor_2019_Exponential,Backdoor_2020_Adversarial,Backdoor_2020_Hash,BD_new_label_2020,Backdoor_2020_New_Label,Backdoor_fragile_2021,Backdoor_2021_Usenix_Entangled,Backdoor_2022_with_Steg,Backdoor_2023_Holo}. Since all the existing works share the same definition of watermark fidelity using learning accuracy, we pay more attention to how the selected trigger set and the corresponding labels are embedded into the host model in these works. We direct the reader to \cite{Survey_2021_NN_Watermark} for a comprehensive and general survey of DNN watermarking.

\begin{table}[!t]
	\renewcommand{\arraystretch}{1}
	\caption{Summary of existing backdoor embedding strategies.}
	\label{BD_embedding}
	\centering
	\begin{spacing}{1}
	\begin{tabular}{c|c|cc}
		\hline
		\hline
		Method & $\#$Trigger/Batch & From Scratch & Fine-Tune \\
		\hline
		2018 \cite{Backdoor_2018_AsiaCCS} & Random & \checkmark & \\
		2018 \cite{Backdoor_2018_Signature} & $50\%$ &&\checkmark  \\
		2018 \cite{Backdoor_2018_turning_weakness}& $2/100$ &\checkmark&  \\
		2019 \cite{Backdoor_2019_Encoder} & $20/120$ &\checkmark& \\
		2019 \cite{Backdoor_2019_Exponential} & $4/100$ &\checkmark& \\
		2020 \cite{Backdoor_2020_Adversarial} & $100\%$ &&\checkmark \\
		2020 \cite{Backdoor_2020_Hash} & $2/100$ &\checkmark& \\
		2020 \cite{BD_new_label_2020} & Random &\checkmark&  \\
		2020 \cite{Backdoor_2020_New_Label} & Random &\checkmark& \\
		2021 \cite{Backdoor_fragile_2021} & Random &&\checkmark  \\
		2021 \cite{Backdoor_2021_Usenix_Entangled} &$\approx 1/3$ &\checkmark & \\
		2022 \cite{Backdoor_2022_with_Steg} & Random & \checkmark &\\
		2023 \cite{Backdoor_2023_Holo} & Random & & \checkmark\\
		\hline 
		\hline
	\end{tabular}
	\end{spacing}
\end{table}

DNN backdoor watermarks can be embedded via either training from scratch or fine-tuning, with mixed training and trigger samples. In \cite{Backdoor_2018_AsiaCCS}, the trigger set is constructed by adding special strings, e.g., "TEST", or noise into a subset of training samples, and backdoor embedding is achieved by training from scratch. In \cite{Backdoor_2018_Signature}, an owner-specific hash based mask is added to training samples for trigger generation, and FTAL is implemented for backdoor embedding. Both embedding from scratch and fine-tuning have been implemented in \cite{Backdoor_2018_turning_weakness}, and embedding from scratch has been found to be the most effective to ensure the accuracies on both the training and trigger data. In \cite{Backdoor_2019_Encoder}, the trigger samples are generated in a more sophisticated way using an encoder taking a training sample and a copyright logo as input, and the embedding strategy is embedding from scratch. The work in \cite{Backdoor_2019_Exponential} is mainly concerned with improving backdoor robustness, while the embedding strategy is also training from scratch. Adversarial examples have been considered in \cite{Backdoor_2020_Adversarial} to generate the trigger set, while in the fine-tuning based embedding process, the original training samples are not used anymore, and the fidelity is controlled by the $50\%$ ``false adversaries'' in the trigger set. In \cite{Backdoor_2020_Hash}, the authors focused on improving backdoor robustness against ambiguity attacks and proposed a one-way-hash-based method to generate the trigger set, which was then embedded via training from scratch. Different from other backdoor based methods, in \cite{BD_new_label_2020}, \cite{Backdoor_2020_New_Label}, and \cite{Backdoor_2022_with_Steg}, the trigger sets generated from frequency, spatial, and steganographic spatial domains, respectively, are assigned to a new label instead of the conventional ``wrong labels''. Because of the introduced extra trigger class, the backdoor can only be embedded through retraining from scratch. Fragile backdoor watermarking is considered in \cite{Backdoor_fragile_2021} to withstand white-box attacks via fine-tuning based embedding. In \cite{Backdoor_2021_Usenix_Entangled}, an entangled watermark embedding from scratch method is proposed to prevent the surrogate attack, based on which the attacker is not able to remove the watermark in surrogate models without significantly sacrificing the functionality. Finally, pseudo-holographic patterns are used as trigger samples in \cite{Backdoor_2023_Holo}.

The above-mentioned backdoor embedding strategies are summarized in Table \ref{BD_embedding}, in which the second column shows how the trigger samples are mixed with normal training samples, and ``Random'' means the batches are randomly chosen from the combined set of trigger and training samples. We observe that the most commonly used embedding method is embedding from scratch, mainly for the consideration of robustness against white-box fine-tuning attacks \cite{Backdoor_2018_turning_weakness}. Although both embedding methods can achieve the fidelity defined by the learning accuracy, the watermarked models actually become different ones from the perspective of deep fidelity, since both the feature extraction and decision boundary of the host model will be altered. Detailed analysis of this is provided in Section \ref{sec_3.2}.

\section{Backdoor Watermark and Fidelity Problem}\label{sec_3}
\subsection{Supervised Learning and Backdoor Embedding}
Consider a supervised learning task for multi-class classification, in which the DNN model is denoted by $M(W,\cdot)$, where $M(\cdot, \cdot)$ describes model architecture and $W$ represents model parameters. The learning set is denoted by the image sample-label pairs $(\mathcal{X}, \mathcal{Y})$, where $\mathcal{Y}=\{0,1,\ldots, K-1\}$, for $K$ classes. The learning process aims to approximate a ground-truth function $f:\mathcal{X} \to \mathcal{Y}$, and it yields a trained model $M(\hat{W}, \cdot)$, where $\hat{W}$ denotes the optimized model parameters, such that\footnote{We follow the protocol that the learning set $\mathcal{X}$ is split into the disjoint training and validation sets, which are used for training and model selection, respectively. Notation-wise, for brevity, we do not explicitly denote the two sets in the equations. Note that ``learning accuracy'' is equivalent to ``validation accuracy'', but the quality of learning is solely determined by the training set.}
\begin{equation}\label{supervised_learning}
	\mathop {\Pr }\limits_{x \in {{\cal X}}} \{ M(\hat{W}, x) \ne f(x)\}  \le \epsilon,
\end{equation} 
where $\epsilon$ is a small positive number. The trained model $M(\hat{W}, \cdot)$ is regarded as the \emph{host model}, which is to be protected via backdoor watermarking.

In backdoor watermarking, the model owner constructs a set of special samples denoted by $\mathcal{T} = \{x_\text{T}^{(1)}, x_\text{T}^{(2)}, \ldots, x_\text{T}^{(n)}\}$, called \emph{trigger set}, and the corresponding  labels are obtained by a function $g: \mathcal{T} \to \mathcal{Y}$, $g\neq f$, denoted by $g(\mathcal{T})$. The tuple $(\mathcal{T}, g(\mathcal{T}))$ is also referred to as the \emph{watermarking key} \cite{Backdoor_2018_turning_weakness}. The design of $(\mathcal{T}, g(\mathcal{T}))$ is the major concern of the existing works \cite{Backdoor_2018_AsiaCCS,Backdoor_2018_Signature,Backdoor_2018_turning_weakness,Backdoor_2019_Encoder,Backdoor_2019_Exponential,Backdoor_2020_Adversarial,Backdoor_2020_Hash,BD_new_label_2020,Backdoor_2020_New_Label,Backdoor_fragile_2021,Backdoor_2021_Usenix_Entangled}. The embedding is successful if the backdoored model $M(\tilde{W}, \cdot)$, obtained by either training from scratch or fine-tuning, satisfies the following conditions,
\begin{subequations}\label{FP_ambiguity_new_label}
	\begin{align}
		\text{Fidelity:} \quad &  \mathop {\Pr } \limits_{x \in {\cal X}\backslash \mathcal{T}} \left\{ {M\big( {\tilde W,x} \big) \ne f(x)} \right\} \le \epsilon,\label{Old_Fidelity}\\
		\text{Backdoor:} \quad & \Pr \limits_{x \in {\cal T}}\left\{ { M\big( {\tilde W, x} \big) \; \ne g(x)} \right\} \le {\epsilon}\label{Backdoor},
	\end{align}
\end{subequations}
where (\ref{Old_Fidelity}) is currently the shared measurement of backdoor fidelity \cite{Backdoor_2018_AsiaCCS,Backdoor_2018_Signature,Backdoor_2018_turning_weakness,Backdoor_2019_Encoder,Backdoor_2019_Exponential,Backdoor_2020_Adversarial,Backdoor_2020_Hash,BD_new_label_2020,Backdoor_2020_New_Label,Backdoor_fragile_2021,Backdoor_2021_Usenix_Entangled}, and (\ref{Backdoor}) ensures the special mapping $g$ contrasting $f$ has been established in the watermarked model. In a closed-loop watermark verification phase, the key $(\mathcal{T}, g(\mathcal{T}))$ can be presented to a judge or an API to verify the consistency between the inference results $M(\tilde{W}, \mathcal{T})$ and the labels $g(\mathcal{T})$ for ownership verification or authentication purposes.

\subsection{Fidelity Problem of Current Approach}\label{sec_3.2}
\begin{figure}[!t]
	\centering
	\subfigure[Host Model.]{\includegraphics[width=2.5in]{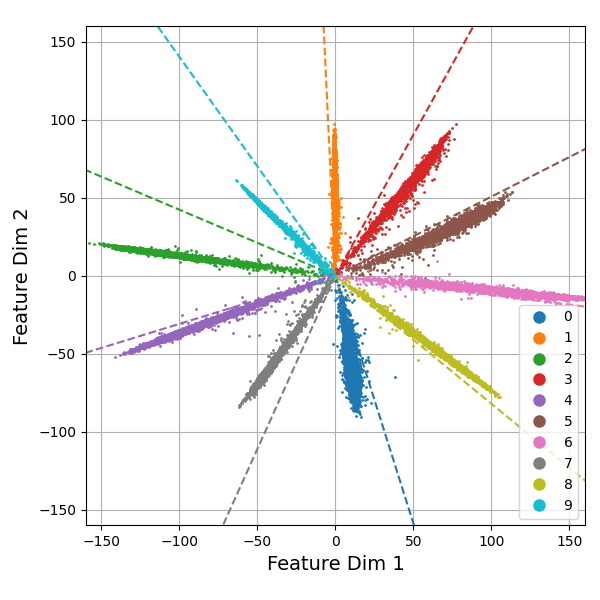}}\quad
	\subfigure[Embedding From Scracth.]{\includegraphics[width=2.5in]{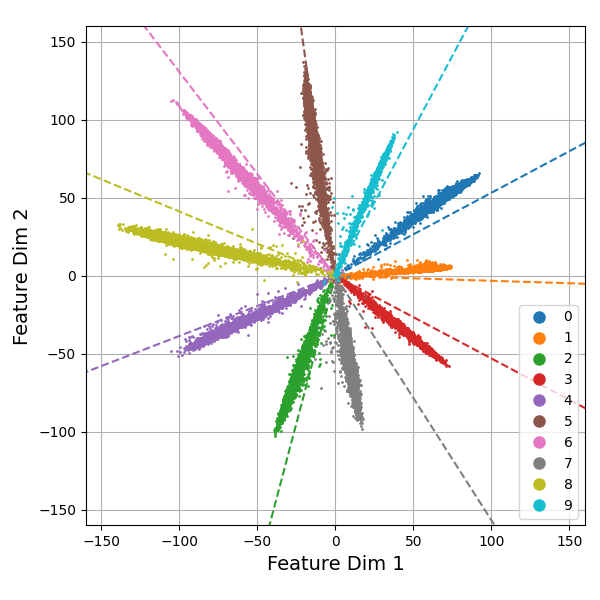}}
	\caption{Demonstration of significantly different feature extraction results on normal training data before and after backdoor embedding, using ResNet18 for MNIST handwritten digit classification, $\mathcal{L}_\text{TWL}=27130.9863$. Dashed lines are the $10$ principal directions of the prototype vectors in the last layer. (a) Host model: validation accuracy $99.47\%$, trigger accuracy $10\%$. (b) Watermarked model: validation accuracy $99.41\%$, trigger accuracy $100\%$. }
	\label{demo_feature}
\end{figure}

Comparing (\ref{Old_Fidelity}) with (\ref{supervised_learning}), it is can be seen that fidelity is considered to be preserved if the two inequalities hold, no matter how the parameters $\tilde{W}$ differ from $\hat{W}$ in the host model. A demonstrative example is presented in Fig. \ref{demo_feature}, in which the number of neurons in the penultimate layer is set to $2$ (with linear activation) for feature visualization. It can be seen that the backdoored model in Fig. \ref{demo_feature} (b) is a high-fidelity model if evaluated by (\ref{Old_Fidelity}), since the learning accuracy only dropped by $0.06\%$ from the host model upon the full embedding of the trigger set. However, because of random initialization and data shuffling in the training from scratch, the resultant functionality of feature extraction is substantially different from that of the host model. For example, in the host model, digit ``$3$'' (red) is likely to be misclassified into the adjacent digit ``$5$'' (brown), but in the watermarked model, ``$3$'' (red) is likely to be misclassified into ``$7$'' (gray) instead. This will result in significant performance differences when the model is deployed in a testing environment in which unseen samples, adversarial examples, and open-set samples widely exist. Therefore, the fidelity defined merely by the learning accuracy is too loose to characterize the inconsistency of the watermarked model over the host model.

Recall the classic imperceptible multimedia watermarking \cite{PR_2022_Watermarking}, regardless of host content format and the measurement metric, watermarking fidelity has always been defined as a type of distance between the watermarked content and the reference host content. Therefore, intuitively, the analogous generic measurement of \emph{embedding distortion} in multimedia watermarking can be interpreted as the total weight loss (TWL) in DNN watermarking, i.e.,
\begin{equation}\label{TWL}
	\mathcal{L}_{\text{TWL}}=\| {\tilde W - \hat W} \|_2^2,
\end{equation}
which is simply the squared $\ell_2$-norm of the differences between the two sets of model parameters viewed as high-dimensional tensors. Nonetheless, it has been pointed out in \cite{Inside_2017_Uchida} and related works that the differences between model parameters become less important compared to model functionalities defined in (\ref{supervised_learning}) and (\ref{Old_Fidelity}) for DNN watermarking. In other words, for models with the same architecture $M(\cdot, \cdot)$, significantly different parameters may still yield an identical learning accuracy, while tiny differences of model parameters may yield significantly different learning accuracies.

Based on the above analysis, neither the learning accuracy in (\ref{Old_Fidelity}) nor the TWL in (\ref{TWL}) can appropriately measure DNN watermarking fidelity, thus it calls for a better approach to tackle this problem. In the following content, we present the proposed concept of deep fidelity as a solution. The TWL will be implemented later in our comparative studies to demonstrate the ineffectiveness of directly incorporating the analogous fidelity concept from multimedia watermarking.

\section{Deep Fidelity}\label{sec_4}
Different from the current fidelity definition in (\ref{Old_Fidelity}) which does not consider the characteristics of the host model and is solely a probabilistic measurement, the proposed deep fidelity is concerned with the two major functionalities of a deep discriminative model for classification tasks, i.e., feature extraction and classification. Therefore, we propose the following interpretation of deep fidelity in the context of image classification.

\emph{\textbf{Deep Fidelity:} Consider a host DNN model for classification tasks, then the watermarked model, upon successful watermark embedding, should preserve both the feature representation of training data and the decision boundary of host model.}

It is worth noting from the above interpretation that first, high-deep-fidelity leads to high-fidelity (\ref{Old_Fidelity}) but not vice versa, and second, deep fidelity is measured using the training data only, because only the training data are used in learning. It is also worth noting that according to a specific model and scenario, deep fidelity can be formulated in different ways (see Section \ref{sec_4.3}), similar to related performance criteria in multimedia watermarking, e.g., imperceptibility can be formulated as peak signal-to-noise ratio, structural similarity and variants, etc., in image watermarking.

\begin{figure}[!t]
	\centering
	\includegraphics[width=5.5in]{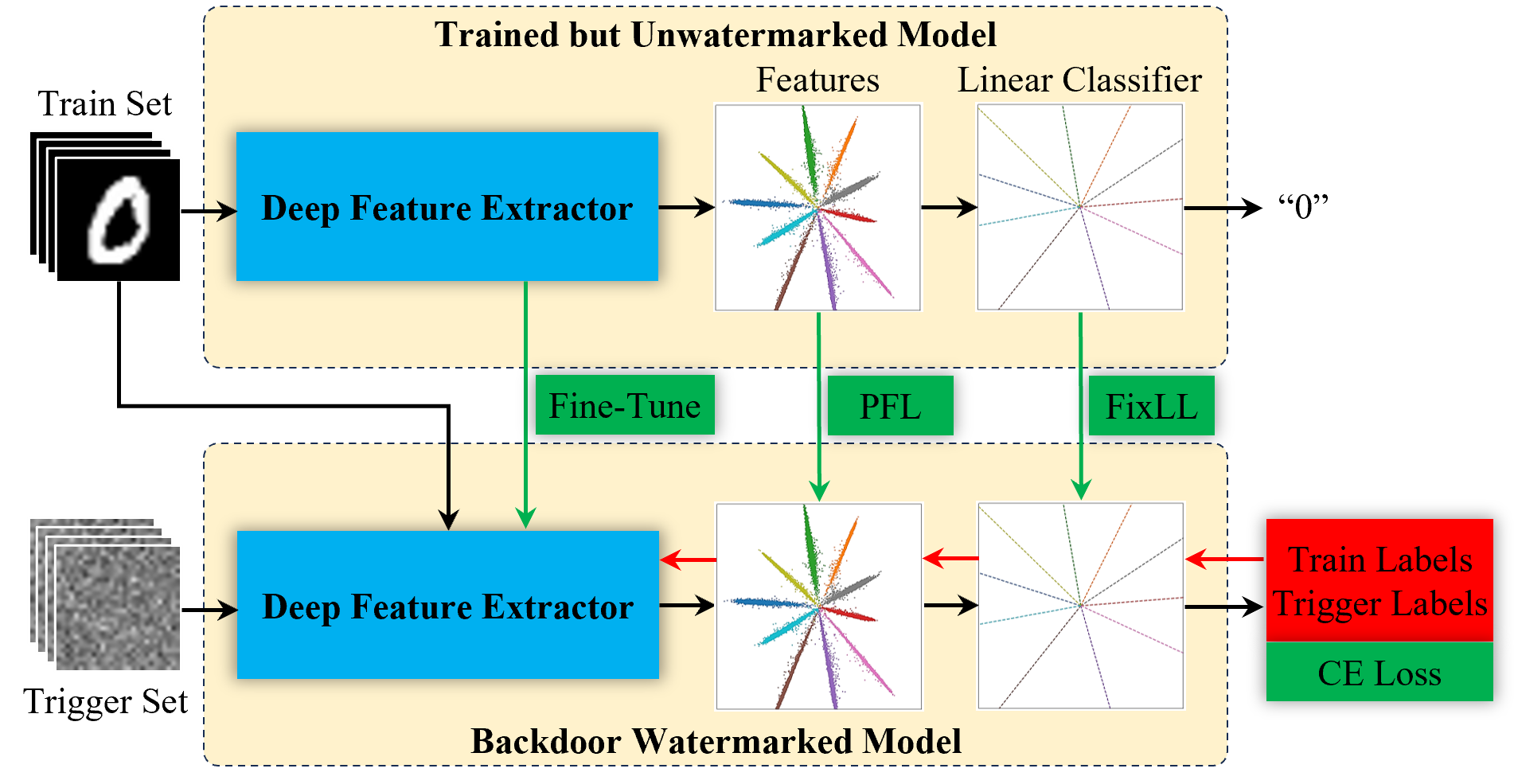}
	\caption{Flowchart of the proposed fine-tuning-based backdoor watermark embedding method to achieve deep fidelity, where MNIST classification and noise pattern trigger samples are considered as examples. The proposed FPL is shown for better visualization, while SPL can also be used. Red components and arrows represent supervision and backpropagation, black arrows represent forward path, while green components and arrows represent treatment and loss functions.}
	\label{flowchart}
\end{figure}

For the host model, let the penultimate layer output feature be $\hat{z}_i\in\mathbb{R}^{d \times 1}$, where $d$ is the feature dimension (the number of neurons in the penultimate layer) and $i$ is the sample index, and let the weights of the last layer classifier be $\hat{W}_\text{P}\in \mathbb{R}^{d \times K}$, where $K$ is the number of classes, then the output of the host model can be expressed as
\begin{equation}
	M(\hat{W}, x_i) = \arg \max \sigma\big( \hat{W}^T_\text{P} \hat{z}_i \big),
\end{equation}
where $\{\cdot\}^T$ is the matrix transpose operator, $\sigma(\cdot)$ is the softmax function, $\hat{z}_i$ is the feature representation of input $x_i$, and $\sigma\big( \hat{W}^T_\text{P} \hat{z}_i \big)\in\mathbb{R}_+^{K\times 1}$ contains the softmax categorical probabilities. Here the bias of the last layer is omitted for brevity \cite{Fix_Pro_2021_TNNLS}. The sign $\hat{\{\cdot\}}$ can be replaced by $\tilde{\{\cdot\}}$ in the above notations to represent the corresponding terms for a watermarked model. We further express the last layer weights as $\hat{W}^T_\text{P} = [w_0, w_1, \ldots, w_{K-1}]$, where $w_i \in \mathbb{R}^{d\times 1}$ is called the prototype vector which determines the categorical direction for the final feature projection. The prototype vectors are known to be perpendicular to the decision boundaries of the final linear classifier, thus they are used as the one-to-one mapped descriptions of the decision boundaries. 

Fig. \ref{flowchart} shows the flowchart of the proposed method to achieve deep fidelity, where MNIST classification is used as an example. Deep fidelity therein is reflected by the minimized difference between the two feature distributions and the unchanged last layer classifier. Note that SPL can replace PFL in the figure, but the PFL is used here for better visualization. Next, we discuss our proposal in detail. For mathematical tractability, we first address the problem of decision boundary, based on which we then present the proposed approaches to preserve the feature representation of training data. 

\begin{figure}[!t]
	\centering
	\includegraphics[width=4.5in]{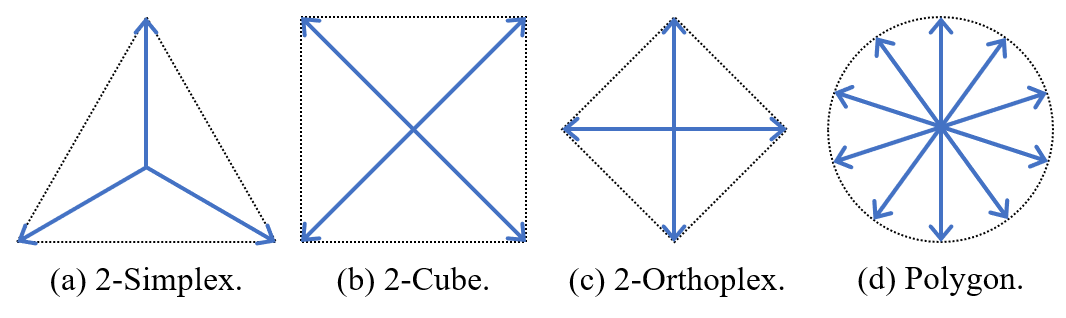}
	\caption{Demonstration of fixed prototype vectors in $\mathbb{R}^{2}$ space ($d=2$) based on regular polytopes \cite{Fix_Pro_2021_TNNLS}, where $K=3$, $4$, $4$, $10$, respectively.}
	\label{Polytope}
\end{figure}

\subsection{Fixed Decision Boundaries}\label{sec_4.1}
Viewing a DNN model for classification as the concatenation of a deep feature extractor and a last layer linear classifier, we propose to fix the last layer (FixLL) during backdoor embedding, expressed by the following constrain
\begin{equation}\label{FixLL}
	\tilde{W}_\text{P} = \hat{W}_\text{P}.
\end{equation}
This setting is inspired by recent discovery on generic deep discriminative models for classification tasks with a fixed classifier \cite{Fix_Pro_2018_ICLR,Fix_Pro_2021_TNNLS}. In the latest and most comprehensive study \cite{Fix_Pro_2021_TNNLS}, it has been rigorously verified that fixing the classifier can not only reduce memory usage without deteriorating classification accuracy, but also improve the quality of feature learning with maximally separable feature representations. This means that assigning fixed values to the prototype weights in the last layer of a DNN model and only training the remaining parameters will cause no performance degradation. Moreover, with a proper setting of the fixed weights, the performance can even be improved, thanks to the prototype-guided maximally separable feature extraction. In \cite{Fix_Pro_2021_TNNLS}, a set of regular polytopes are proposed to design the fixed weights, and the $2$-dimensional instances are provided in Figs. \ref{Polytope} (a), (b), and (c), respectively. Situations for higher dimensional spaces vary and can be found in \cite{Fix_Pro_2021_TNNLS}. Besides, the maximally and uniformly separable weight assignment for $2$-dimensional feature extraction and $10$-class classification is demonstrated in Fig. \ref{Polytope} (d). Comparing Fig. \ref{Polytope} (d) and Fig. \ref{demo_feature} (a), it can be seen that training with learnable decision boundaries eventually tends to approximate the uniform separation of the feature space. Note that FixLL can work with both embedding from scratch and embedding via fine-tuning, while in the latter scheme, FixLL can be imposed either during the fine-tuning or in the original training stage.

\subsection{Preserving Feature Representation}
Based on the fixed classifier, we now address the other part of deep fidelity, i.e., the task of preserving the mechanism for training data feature extraction. Ideally, we want $\forall i\in |\mathcal{X}|$, $\tilde{z}_i = \hat{z}_i$, but given the fixed classifier, this may lead to the trivial case that $M(\tilde{W}, \cdot)= M(\hat{W}, \cdot)$, in which the watermarks cannot be embedded. Therefore, the preservation of feature representation can be cast as an optimization problem of embedding the backdoor with the minimum statistical alteration from $\hat{z}_i$ to $\tilde{z}_i$. Accordingly, we propose the penultimate feature loss (PFL),
\begin{equation}\label{PFL}
	\mathcal{L}_{\text{PFL}|\mathcal{X}}=\sum\limits_{i} \|{\tilde{z}_i - \hat{z}_i\|_2^2},
\end{equation}
which is the sum of the squared distances between features extracted by the watermarked and host models from the same sample. Compared to the TWL in (\ref{TWL}), the proposed PFL is more descriptive in that it directly characterizes how the features extracted by the watermarked model deviate from those extracted by the host model. 

Alternatively, since $\tilde{W}_\text{P} = \hat{W}_\text{P}$ is a constant during watermark embedding, the softmax probabilities can also be used to indirectly measure the difference between feature representations, which leads to the proposed softmax probability-distribution loss (SPL),
\begin{align}
	\mathcal{L}_{\text{SPL}|\mathcal{X}} & = \sum\limits_{i} \mathcal{D}_{\text{KL}}\left[\sigma\big(\hat{W}_\text{P}^T\hat{z}_i\big)\Big|\Big|\sigma\big(\tilde{W}_\text{P}^T\tilde{z}_i\big)\Big.\Big.\right]\notag\\
	& = \sum\limits_{i} \mathcal{D}_{\text{KL}}\left[\sigma\big(\hat{W}_\text{P}^T\hat{z}_i\big)\Big|\Big|\sigma\big(\hat{W}_\text{P}^T\tilde{z}_i\big)\Big.\Big.\right],\label{SPL}
\end{align}
where $\mathcal{D}_{\text{KL}}(\cdot ||\cdot)$ calculates the Kullback-Leibler divergence (KLD) between two probability distributions, and the second equality is based on (\ref{FixLL}). The minimization of the SPL pushes the categorical probabilities $\sigma\big(\tilde{W}_\text{P}^T\tilde{z}_i\big)$ towards $\sigma\big(\hat{W}_\text{P}^T\hat{z}_i\big)$, thus the feature distribution can be preserved. Note that both the PFL and SPL are a function of $\tilde{z}_i$, and they both become zero if $\forall i \in \{0, 1, \ldots, |\mathcal{X}|-1\}$, $\tilde{z}_i=\hat{z}_i$. However, the geometric surfaces of the two loss functions are different, which can lead to different optimization results. Also note that both PFL and SPL are conditioned on the training data only, in which the trigger set is not involved.

\subsection{Watermark Embedding}\label{sec_4.3}
We now present several high-deep-fidelity backdoor watermark embedding schemes incorporating the above analysis and designs. For comparison purpose, the existing FTLL and FTAL embedding methods are also summarized here in a unified view. Given the training set $(\mathcal{X}, \mathcal{Y})$, the watermarking key $(\mathcal{T}, g(\mathcal{T}))$, all fine-tuning based methods are initialized by setting $\tilde{W}=\hat{W}$, and the existing and proposed backdoor embedding methods are derived as follows.

\subsubsection{Fine-Tune Last Layer (FTLL) \texorpdfstring{\cite{Backdoor_2018_turning_weakness}}{}}
From the viewpoint of deep fidelity, the FTLL method does not alter the way of feature extraction, while only the last layer classifier $\tilde{W}_\text{P}$ is fine-tuned for backdoor embedding. Therefore, FTLL can be expressed as
\begin{equation}\label{FTLL_Embedding}
	\tilde{W}_\text{P} =  \arg \mathop {\min }\limits_{\tilde{W}_\text{P}} \mathcal{L}_{\rm{CE}}\left[{{\cal X}} \cup {{\cal T}}, f(\mathcal{X})\cup g(\mathcal{T})\right],
\end{equation}
where $\mathcal{L}_\text{CE}(\cdot, \cdot)$ is the cross-entropy loss. The training set and trigger set are combined as a single dataset in the fine-tuning process, in which the former prevents the model from forgetting the training data, and the latter is used for the model to learn the backdoor. The FTLL methods can perfectly preserve feature extraction, but the features of the trigger samples are also fixed, which may not be separable via only tuning the classifier. Therefore, this method is not able to reliably embed the backdoor into the host model.  

\subsubsection{Fine-Tune All Layers (FTAL)  \texorpdfstring{\cite{Backdoor_2018_Signature,Backdoor_2018_turning_weakness,Backdoor_2020_Adversarial,Backdoor_fragile_2021}}{}}
The FTAL method provides more degrees of freedom for the model to learn the backdoor. Instead of only updating $\tilde{W}_\text{P}$, FTAL updates all parameters, which is given by
\begin{equation}\label{FTAL_Embedding}
	\tilde W = \arg \mathop {\min }\limits_{\tilde{W}} \mathcal{L}_{\rm{CE}}\left[{{\cal X}} \cup {{\cal T}}, f(\mathcal{X})\cup g(\mathcal{T})\right].
\end{equation}
Although the alteration of model parameters can be indirectly controlled by using a very small learning rate, this method lacks the active control of both training data feature extraction and decision boundaries. Note that embedding from scratch shares the same optimization in (\ref{FTAL_Embedding}), while it only differs by unlearned parameter initialization.  

\subsubsection{FTAL\texorpdfstring{$+$}{}TWL}
An intuitive treatment to improve the control of FTAL is to add the TWL in (\ref{TWL}) during watermark embedding, as discussed in Section \ref{sec_3.2}, and the corresponding embedding method is given by
\begin{equation}\label{FTAL+TWL_Embedding}
	\tilde W = \arg \mathop {\min }\limits_{\tilde{W}} \mathcal{L}_{\rm{CE}}\left[{{\cal X}} \cup {{\cal T}}, f(\mathcal{X})\cup g(\mathcal{T})\right] + {\alpha}\mathcal{L}_{\text{TWL}},
\end{equation}
where $\alpha$ is a weighting hyperparameter. The TWL in (\ref{FTAL+TWL_Embedding}) serves as a regularizer to ensure $\tilde{W}$ does not deviate from $\hat{W}$ in an overall scale. This regularizer is analogous to the embedding distortion in multimedia watermarking, but it will later be shown to be ineffective in DNN watermarking.

\subsubsection{Fix Last Layer (FixLL) \texorpdfstring{\cite{Fix_Pro_2021_TNNLS}}{}}
An alternative approach to control FTAL is to incorporate the results in \cite{Fix_Pro_2018_ICLR,Fix_Pro_2021_TNNLS}, i.e., directly fixing the last layer, as discussed in Section \ref{sec_4.1}. This method is in fact a counterpart to FTLL, in that the fixed and updated portions of parameters are swapped in the two methods. The FixLL method can then be expressed as
\begin{align}
	\tilde W = & \arg \mathop {\min }\limits_{\tilde{W}} \mathcal{L}_{\rm{CE}}\left[{{\cal X}} \cup {{\cal T}}, f(\mathcal{X})\cup g(\mathcal{T})\right]\label{FixLL_Embedding}\\
	& \text{ subject to}\quad \tilde{W}_\text{P} = \hat{W}_\text{P}\notag.
\end{align}
The feature extraction portion of a DNN model is known to be more powerful than the last layer classifier \cite{Fix_Pro_2021_TNNLS}, thus FixLL is expected to be more effective than FTLL in terms of watermark embedding. FixLL can work for both embedding from scratch and embedding via fine-tuning. 

\subsubsection{FixL\texorpdfstring{$+$}{}TWL}
The TWL can be added to the FixLL method to indirectly control feature extraction, which leads to 
\begin{align}
	\tilde W = &\arg \mathop {\min }\limits_{\tilde{W}} \mathcal{L}_{\rm{CE}}\left[{{\cal X}} \cup {{\cal T}}, f(\mathcal{X})\cup g(\mathcal{T})\right] + {\alpha}\mathcal{L}_{\text{TWL}} \label{FixLL+TWL_Embedding} \\
	& \text{ subject to}\quad \tilde{W}_\text{P} = \hat{W}_\text{P}\notag.
\end{align}
It only differs from FTAL$+$TWL by the additional constraint.

\subsubsection{FixL\texorpdfstring{$+$}{}penultimate feature loss (PFL)}
The proposed feature representation losses can be used with FixLL to achieve active control of both feature extraction and decision boundaries. With the use of the PFL, the proposed embedding method can be expressed as
\begin{align}
	\tilde W = &\arg \mathop {\min }\limits_{\tilde{W}} \mathcal{L}_{\rm{CE}}\left[{{\cal T}},  g(\mathcal{T})\right] + {\beta}\mathcal{L}_{\text{PFL}|\mathcal{X}} \label{FixLL+PFL} \\
	& \text{ subject to}\quad \tilde{W}_\text{P} = \hat{W}_\text{P}\notag,
\end{align}
where $\beta$ is a weighting hyperparameter. The FixLL$+$PFL method differs from all the previous methods (\ref{FTLL_Embedding})\---(\ref{FixLL+TWL_Embedding}) in that it does not combine the training and trigger data. Instead, the trigger samples are learned through cross-entropy minimization, while the PFL, which is conditioned on the training set in $\mathcal{X}$, is used to ensure that the model not only does not forget the original training data, but also preserves the feature representation of the host model.

\begin{table}[!t]
	\renewcommand{\arraystretch}{1.}
	\setlength{\tabcolsep}{3pt}
	\caption{Qualitative comparison of backdoor embedding methods.}
	\label{Comparison}
	\centering
	\begin{tabular}{c|c|c|c|c|c}
		\hline
		\hline
		&\multirow{3}{*}{Method} & \multirow{3}{*}{\begin{tabular}{@{}c@{}}Weight\\Loss\end{tabular}} & \multicolumn{3}{c}{Sub-Task}\\
		\cline{4-6}
		&&& Trigger & Feature & Decision \\
		&&& Learning & Representation & Boundary\\
		\hline
		\multirow{2}{*}{\textbf{ A }}&FTLL &&&\checkmark&\\
		&FTAL &&\checkmark&&\\
		\hline
		\multirow{3}{*}{\textbf{ B }}&FTAL$+$TWL &\checkmark&\checkmark&&\\
		&FixLL &&\checkmark&&\checkmark\\
		&FixLL$+$TWL &\checkmark&\checkmark&&\checkmark\\
		\hline
		\multirow{2}{*}{\textbf{ C }}&FixLL$+$PFL &&\checkmark&\checkmark&\checkmark\\
		&FixLL$+$SPL &&\checkmark&\checkmark&\checkmark\\
		\hline
		\hline
	\end{tabular}
\end{table}

\subsubsection{FixLL\texorpdfstring{$+$}{}Softmax Probability-Distribution Loss (SPL)}
Replacing the PFL in (\ref{FixLL+PFL}) by the SPL, the proposed FixLL$+$SPL method is given by
\begin{align}
	\tilde W = & \arg \mathop {\min }\limits_{\tilde{W}} \mathcal{L}_{\rm{CE}}\left[{{\cal T}}, g(\mathcal{T})\right] + {\gamma}\mathcal{L}_{\text{SPL}|\mathcal{X}} \label{FixLL+SPL},\\
	& \text{ subject to}\quad \tilde{W}_\text{P} = \hat{W}_\text{P}\notag,
\end{align}
where $\gamma$ is another weighting hyperparameter, and the FixLL constraint is not needed since it is satisfied in $\mathcal{L}_{\text{SPL}|\mathcal{X}}$ (see (\ref{SPL})). In the proposed two embedding methods, $\mathcal{L}_{\rm{CE}}\left[{{\cal T}}, g(\mathcal{T})\right]$ ensures successful backdoor embedding, the PFL or SPL preserves feature representation, while the constraint (\ref{FixLL}) ensures unaltered decision boundaries. The qualitative comparison among all the embedding strategies presented in this subsection is summarized in Table \ref{Comparison}, where Group \textbf{A} contains the existing embedding methods \cite{Backdoor_2018_Signature,Backdoor_2018_turning_weakness,Backdoor_2020_Adversarial,Backdoor_fragile_2021}, Group \textbf{B} are the direct implementations of FixLL \cite{Fix_Pro_2021_TNNLS} and multimedia-watermarking-analogous methods, while Group \textbf{C} contains the proposed methods. It can be seen from Table \ref{Comparison} that the three sub-tasks during watermark embedding have been disentangled and tackled respectively via our proposal. Therefore, both FixLL$+$PFL and FixLL$+$SPL can achieve deep fidelity.

\section{Analysis and Experiments}\label{sec_5}
In this section, we comprehensively analyze the performances of the proposed high-deep-fidelity DNN backdoor watermarking methods in comparison with the multimedia-watermarking-analogous methods as well as the existing methods \cite{Backdoor_2018_Signature,Backdoor_2018_turning_weakness,Backdoor_2020_Adversarial,Backdoor_fragile_2021,Backdoor_2023_Holo}. Since advanced model design and superior classification performance are not of our interest, we use several standard models, e.g., ResNet \cite{ResNet} and WRN28\_10, for the analysis and experiments. Specifically, a lightweight modification of ResNet18 is used to perform classification tasks on MNIST and CIFAR-10 datasets. The model structure is summarized in Table \ref{model}, where the fully-connected penultimate layer has $2$, $3$, and $64$ neurons, respectively. The versions of $2$ and $3$ penultimate layer neurons are used for visualization purpose, while they can also exhibit promising results. To verify the generalizability of the results on more complicated tasks, the WRN28\_10 model for CIFAR-100 classification task is also considered as the host model in the experiments.

\begin{table}[!t]
	\renewcommand{\arraystretch}{1}
	\caption{Architecture of ResNet18 used in this paper.}
	\label{model}
	\centering
	\begin{tabular}{c|c}
		\hline
		\hline
		Layer & Structure\\
		\hline
		&  $7\times 7$, $16$, stride $1$, $3\times 3$ maxpool, stride $2$\\ 
		\hline
		layer1 & $[\text{Basic Block},16] \times 2$\\
		layer2 & $[\text{Basic Block},32] \times 2$\\
		layer3 & $[\text{Basic Block},64] \times 2$\\
		layer4 & $[\text{Basic Block},128] \times 2$\\
		\hline
		& global average pooling\\
		\hline
		penultimate & $d=\{2, 3, 64\}$ \\
		\hline
		output & linear $10$, softmax\\
		\hline
		\hline
	\end{tabular}
\end{table}

\emph{Hyperparameter Setting:} For the training of the host model, we use the Adam optimizer with the default learning rate of $0.001$, while during fine-tuning based backdoor embedding, the learning rate is reduced to $0.0001$, following the setting in \cite{Backdoor_2020_Adversarial}. The maximum number of epochs is set to $100$, while the learning will terminate if the objectives of training or backdoor embedding can be satisfied earlier. Batch size is set to $32$ as a constant across all experiments, while for each batch, a few randomly selected trigger samples are appended to the batch for backdoor embedding. The trigger set contains random noise patterns under the normal distribution, while similar type of trigger samples have been considered in the existing works \cite{Backdoor_2018_AsiaCCS,Backdoor_2020_Adversarial,Backdoor_2020_Hash,Backdoor_fragile_2021}. The model with the best validation accuracy is selected as host model, while validation accuracy, trigger accuracy, and deep fidelity are jointly considered for the selection of backdoored models.

\emph{Performance Metrics:} We use validation accuracy (Val acc) to measure the model's original functionality, trigger accuracy (Trig acc) to measure the backdoor watermark verification result, $\mathcal{L}_\text{TWL}$ to measure the alteration of model weights after backdoor embedding, average $\mathcal{L}_{\text{PFL}|\mathcal{X}}$ and average $\mathcal{L}_{\text{SPL}|\mathcal{X}}$ to measure the preservation of feature extraction mechanism (training data feature distribution), and $\|\hat{W}_\text{P} - \tilde{W}_\text{P} \|^2_2$ for the alteration of decision boundary. 

\begin{figure}[!t]
	\centering
	\subfigure[Host Prototype.]{\includegraphics[width=1.5in]{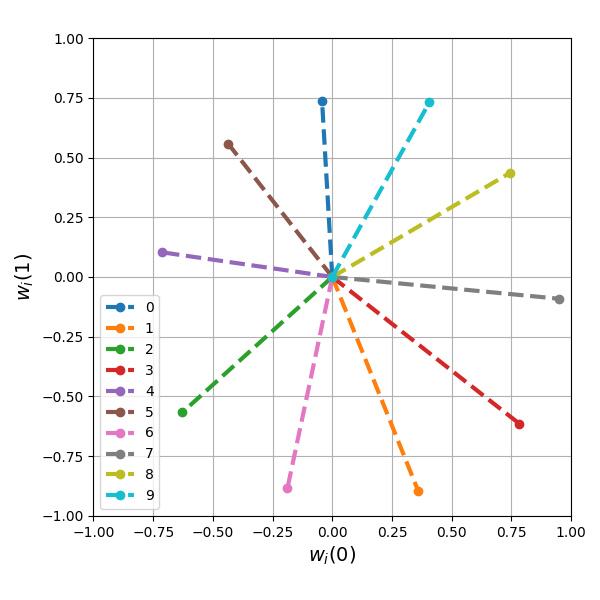}}
	\subfigure[FTLL Prototype.]{\includegraphics[width=1.5in]{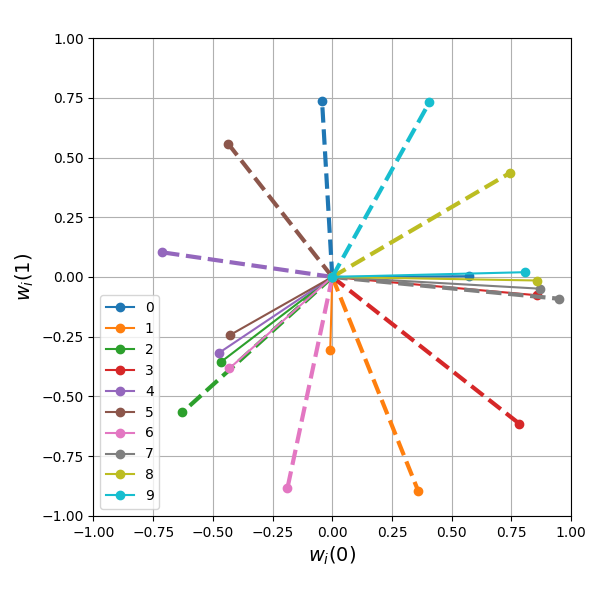}}
	\subfigure[FTAL Prototype.]{\includegraphics[width=1.5in]{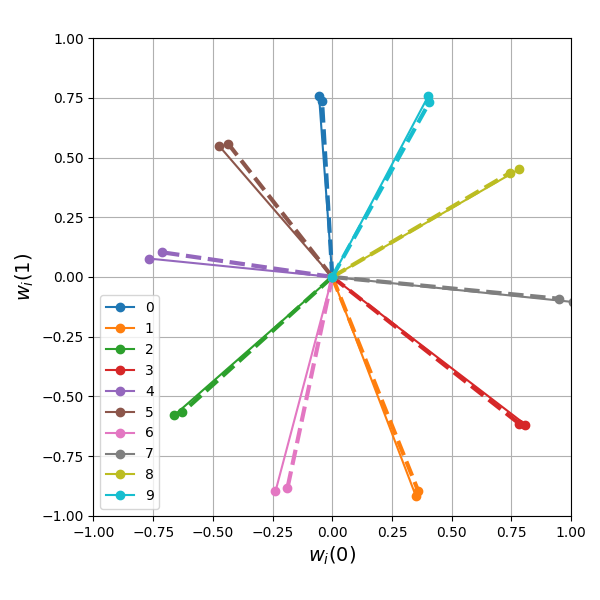}}
	\subfigure[FTAL+TWL Prototype.]{\includegraphics[width=1.5in]{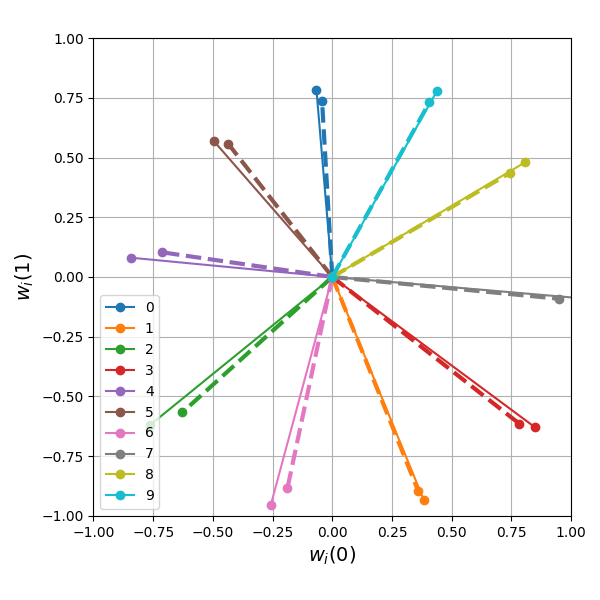}}
	\hfill\\
	\vspace{-6pt}
	\subfigure[Host.]{\includegraphics[width=1.5in]{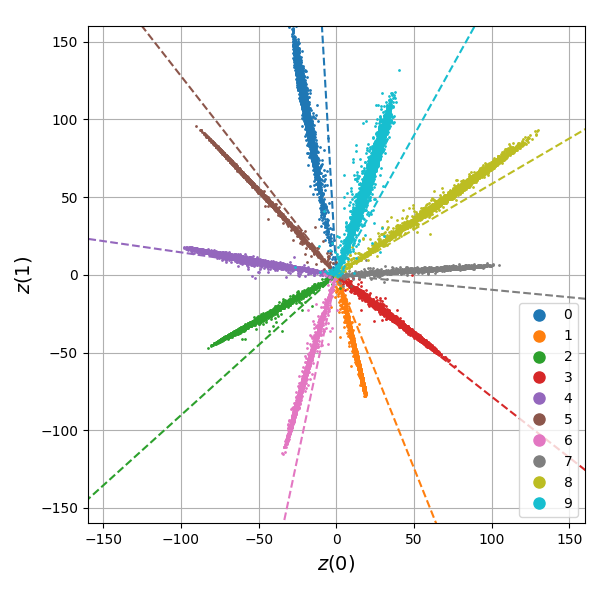}}
	\subfigure[FTLL.]{\includegraphics[width=1.5in]{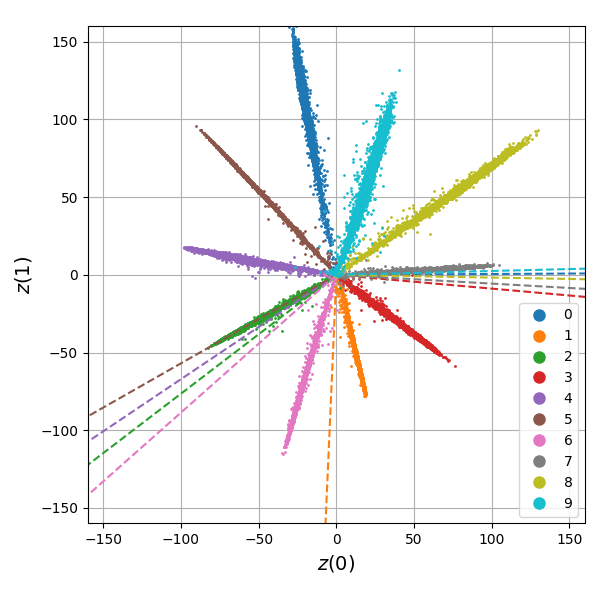}}
	\subfigure[FTAL.]{\includegraphics[width=1.5in]{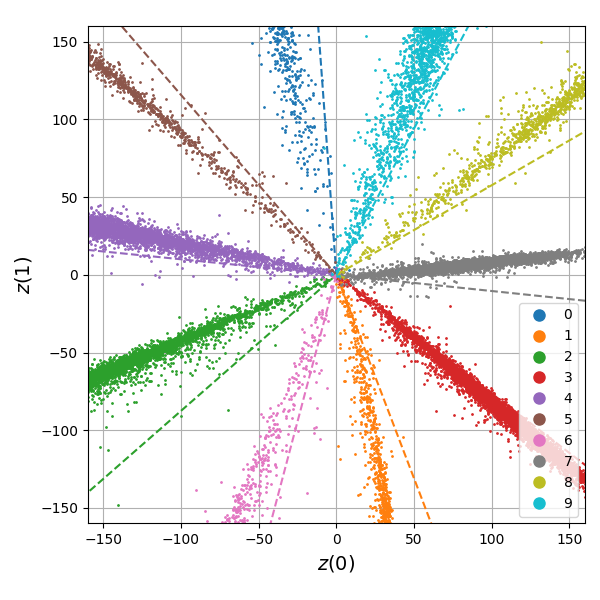}}
	\subfigure[FTAL$+$TWL.]{\includegraphics[width=1.5in]{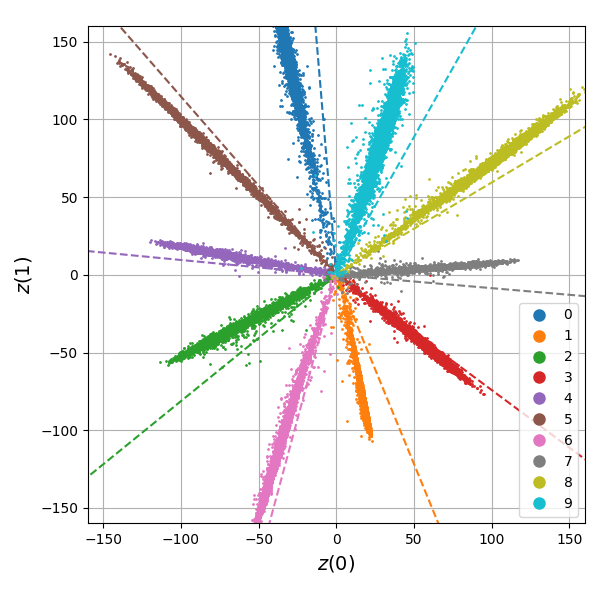}}
	\hfill\\
	\vspace{-6pt}
	\subfigure[FixLL.]{\includegraphics[width=1.5in]{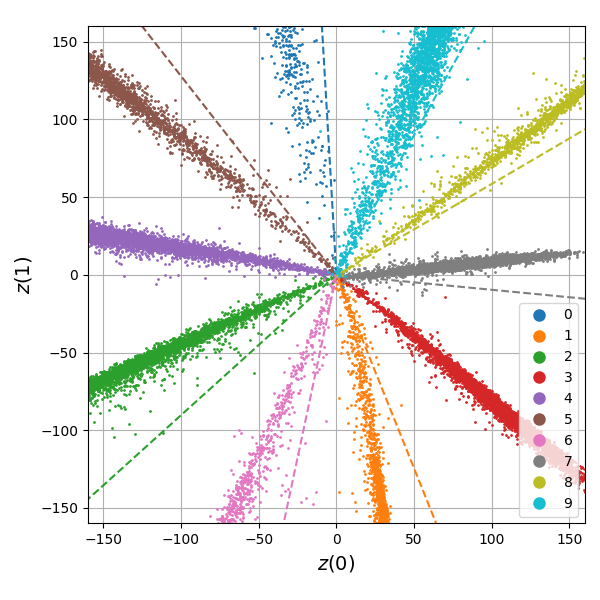}}
	\subfigure[FixLL$+$TWL.]{\includegraphics[width=1.5in]{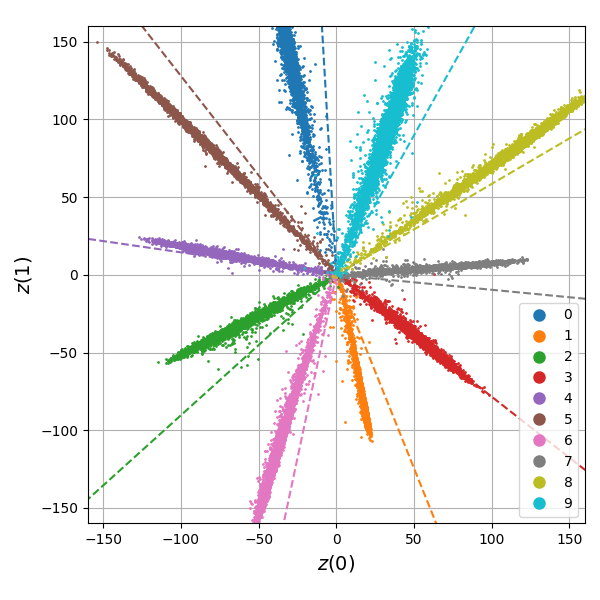}}
	\subfigure[FixLL$+$PFL.]{\includegraphics[width=1.5in]{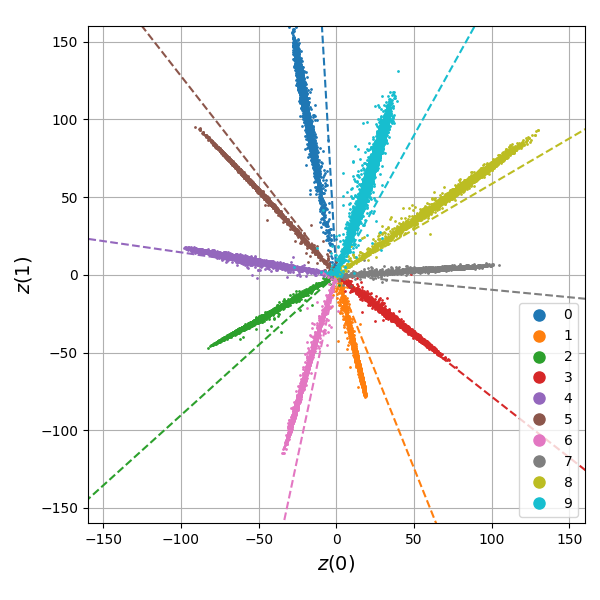}}
	\subfigure[FixLL$+$SPL.]{\includegraphics[width=1.5in]{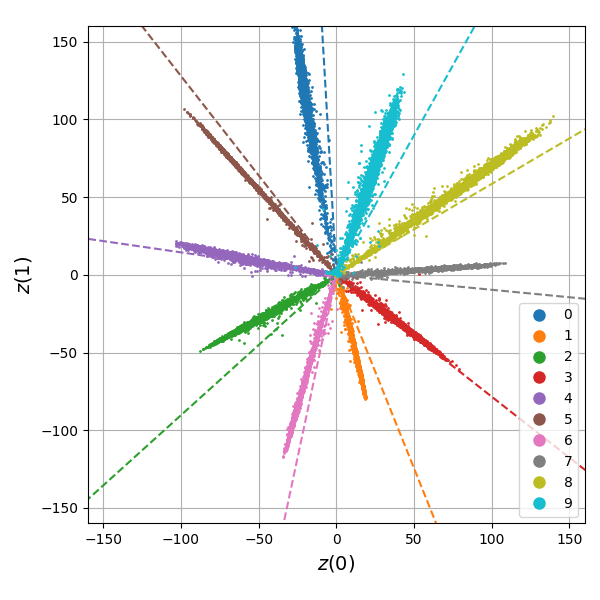}}
	\hfill\\
	\caption{Visualization of deep fidelity in terms of training data feature representation (e)\---(l) and decision boundary (a)\---(d) using MNIST dataset, where $d=2$ ($2$-dimensional visualization), $m=4$, $\alpha=\beta=0.01$, $\gamma=1000$. In (a)\---(d), dashed lines are the prototype vectors of the host model. Prototype vectors of FixLL-based methods (i)\---(\l) are identical to those of the host model in (a) and are not plotted for brevity. Only FTLL (b) (e) failed backdoor embedding.}
	\label{visualization_2D}
\end{figure}

\begin{figure}[!t]
	\centering
	\subfigure[Host Prototype.]{\includegraphics[width=1.5in]{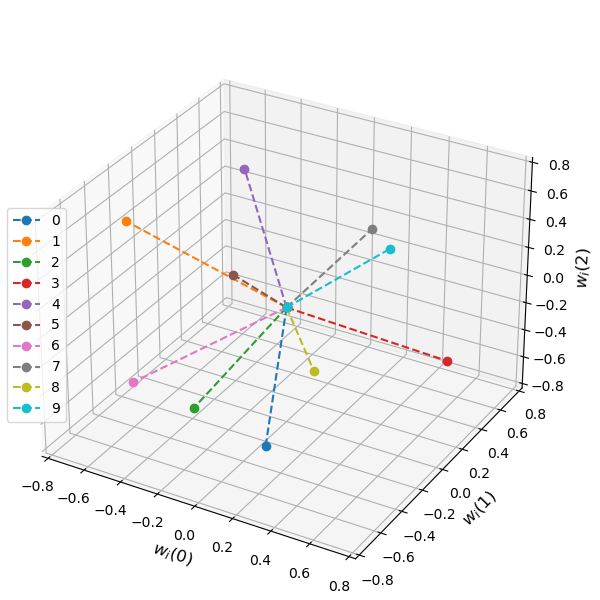}}
	\subfigure[FTLL Prototype.]{\includegraphics[width=1.5in]{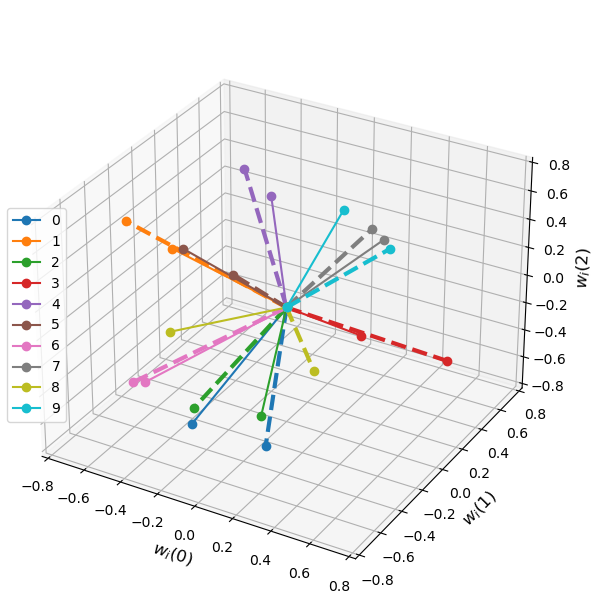}}
	\subfigure[FTAL Prototype.]{\includegraphics[width=1.5in]{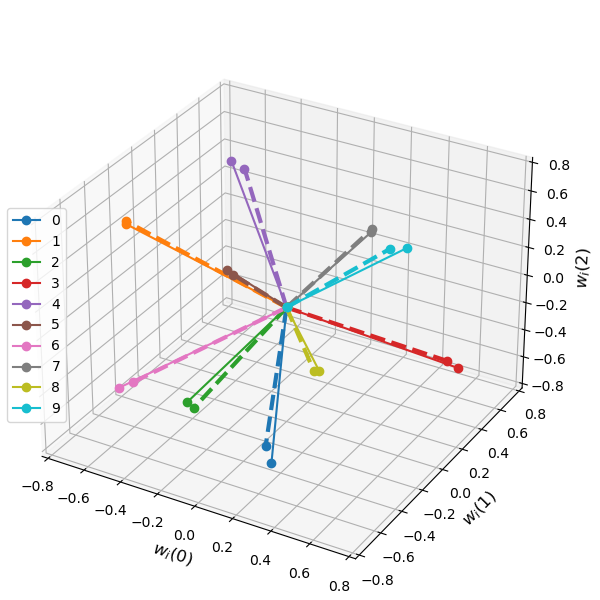}}
	\subfigure[FTAL+TWL Prototype.]{\includegraphics[width=1.5in]{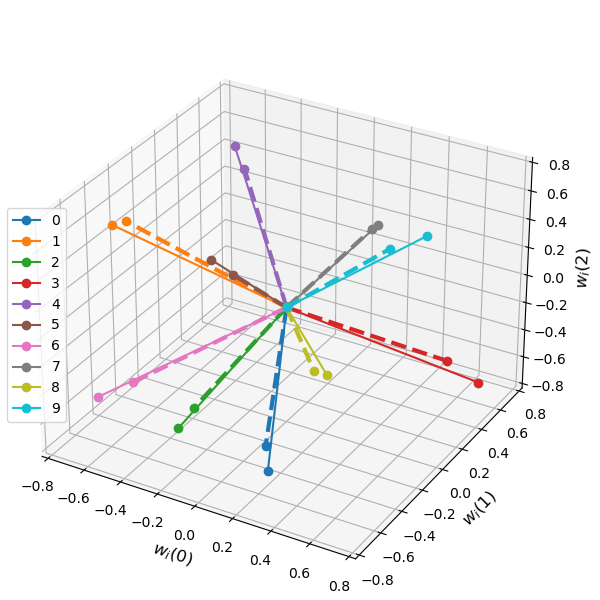}}
	\hfill\\
	\vspace{-10pt}
	\subfigure[Host.]{\includegraphics[width=1.5in]{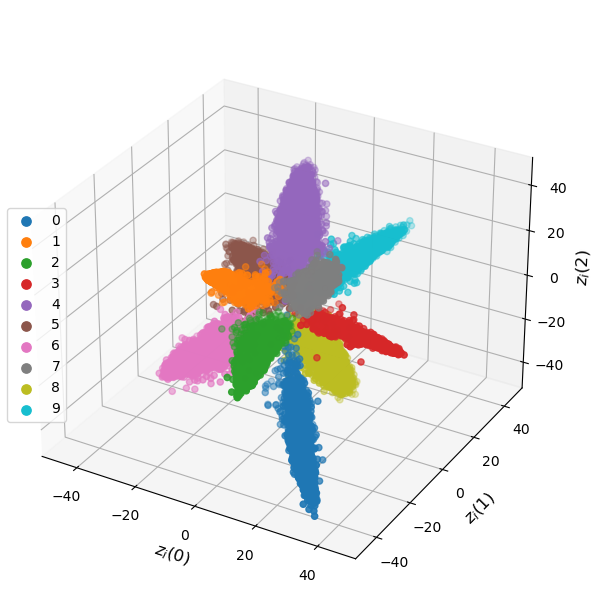}}
	\subfigure[FTLL.]{\includegraphics[width=1.5in]{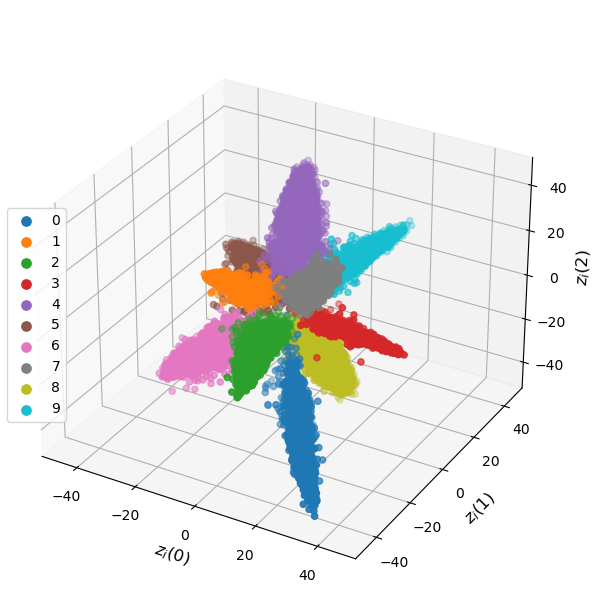}}
	\subfigure[FTAL.]{\includegraphics[width=1.5in]{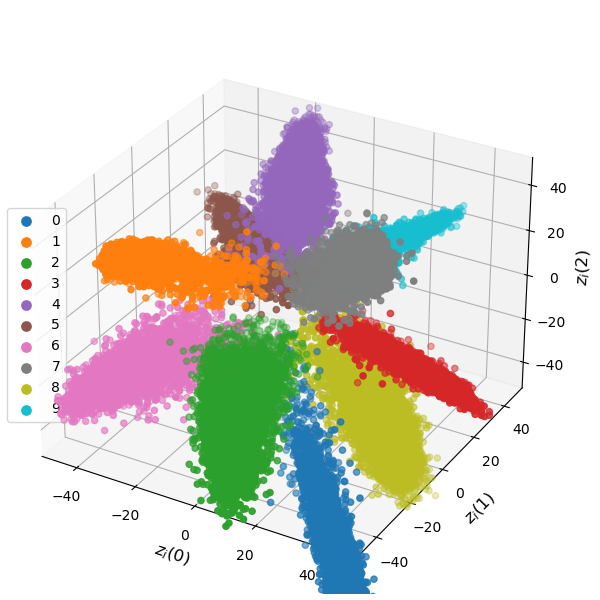}}
	\subfigure[FTAL$+$TWL.]{\includegraphics[width=1.5in]{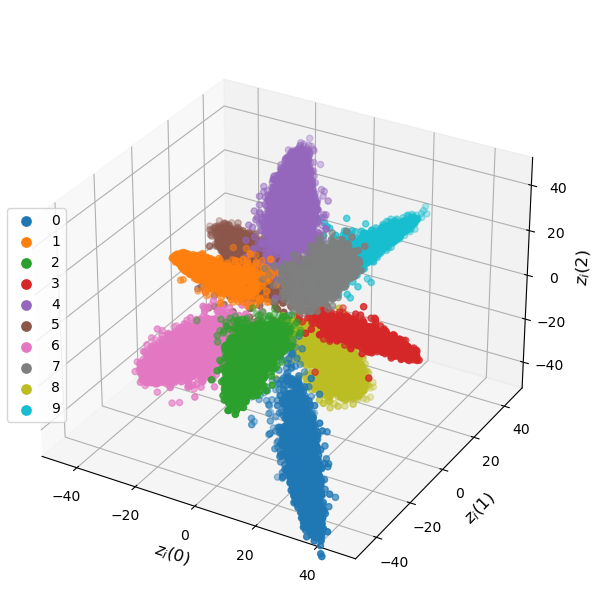}}
	\hfill\\
	\vspace{-10pt}
	\subfigure[FixLL.]{\includegraphics[width=1.5in]{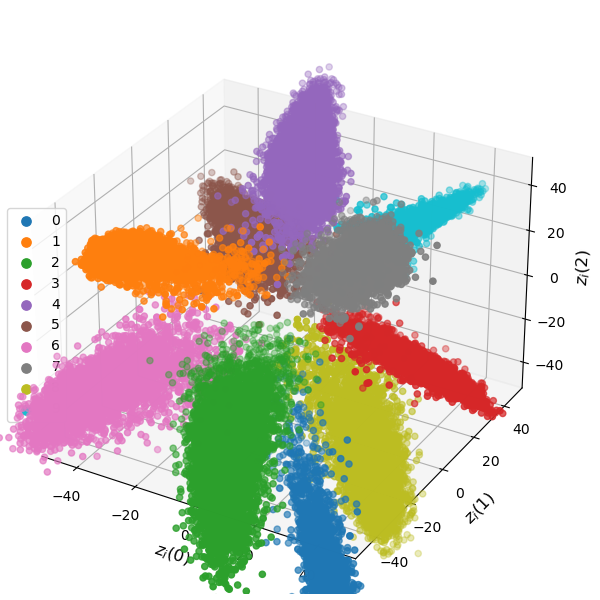}}
	\subfigure[FixLL$+$TWL.]{\includegraphics[width=1.5in]{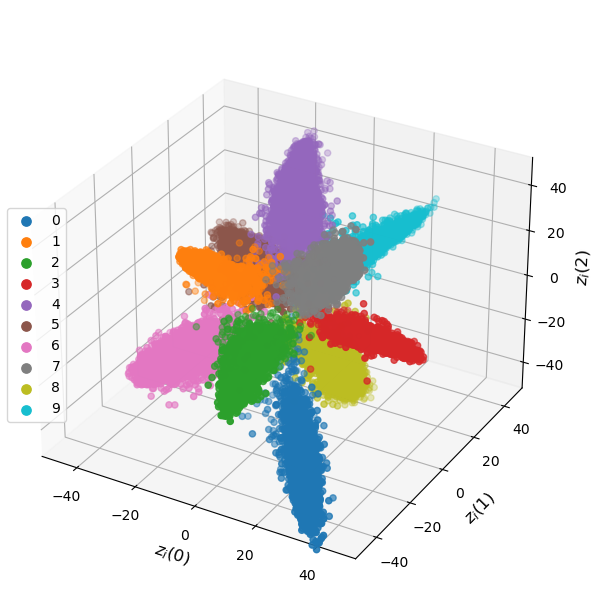}}
	\subfigure[FixLL$+$PFL.]{\includegraphics[width=1.5in]{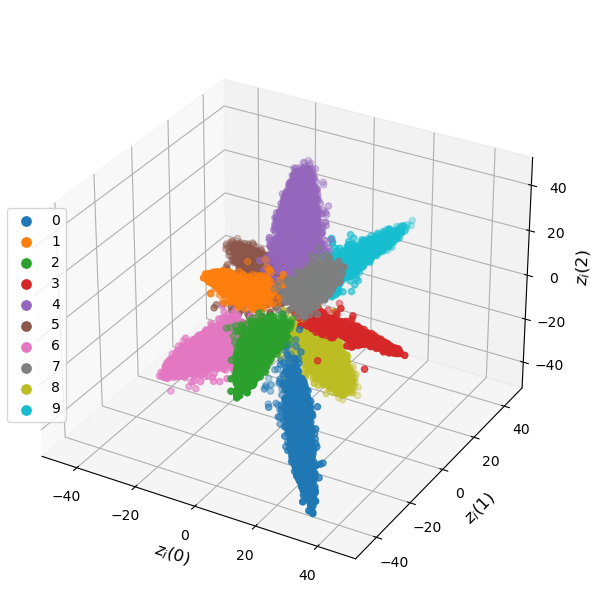}}
	\subfigure[FixLL$+$SPL.]{\includegraphics[width=1.5in]{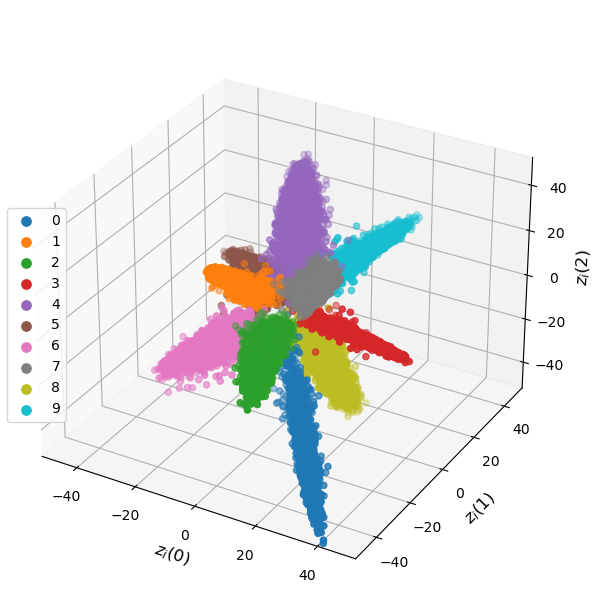}}
	\hfill\\
	\caption{Visualization of deep fidelity in terms of training data feature representation (e)\---(l) and decision boundary (a)\---(d) using MNIST dataset, where $d=3$ ($3$-dimensional visualization), $m=4$, $\alpha=\beta=0.01$, $\gamma=1000$. In (a)\---(d), dashed lines are the prototype vectors of the host model. Prototype vectors of FixLL-based methods (i)\---(\l) are identical to those of the host model in (a) and are not plotted for brevity. Only FTLL (b) (e) failed backdoor embedding.}
	\label{visualization_3D}
\end{figure}

\subsection{Deep Fidelity Visualization on MNIST}
The $2$- and $3$-dimensional visualization results of MNIST training data feature distributions and decision boundaries of the backdoor embedding methods (\ref{FTLL_Embedding})\---(\ref{FixLL+SPL}) are presented in Figs. \ref{visualization_2D} and \ref{visualization_3D}, where $100$ noise pattern trigger samples with random labels are used for backdoor embedding, and each class is consistently represented by a unique color. The corresponding quantitative measurements are summarized in Table \ref{quantitative}. The first rows in Figs. \ref{visualization_2D} and \ref{visualization_3D} compare the prototype vectors in $\hat{W}_\text{P}$ and $\tilde{W}_\text{P}$. When $d=2$ and $K=10$, we can express the vectors as 
\begin{equation}\label{w0w1}
	\left[ {\left[ {\begin{array}{*{20}{c}}
				{{w_0}(0)} \\ 
				{{w_0}(1)} 
		\end{array}} \right],\left[ {\begin{array}{*{20}{c}}
				{{w_1}(0)} \\ 
				{{w_1}(1)} 
		\end{array}} \right], \ldots ,\left[ {\begin{array}{*{20}{c}}
				{{w_9}(0)} \\ 
				{{w_9}(1)} 
		\end{array}} \right]} \right],
\end{equation}
for both models, which are used as the axes of Figs. \ref{visualization_2D} (a)\---(d). Similarly, axes of Figs. \ref{visualization_2D} (e)\---(l) are the respective dimension of the $2$-dimensional features. In Figs. \ref{visualization_2D} (e)\---(l), the dashed lines show the directions of the prototype vectors. Identical settings apply to Fig. \ref{visualization_3D}, while (\ref{w0w1}) can be extended to the case of $d=3$ to determine the axes in Fig. \ref{visualization_3D}. According to Table \ref{quantitative} and Fig. \ref{visualization_2D}, the performances of the implemented methods are discussed as follows. Additional remarks including the coverage of results in Fig. \ref{visualization_3D} will be provided subsequently.

\textbf{FTLL.} It can be seen from Fig. \ref{visualization_2D} (b) that the prototype vectors after FTLL (solid lines) are significantly altered from the reference model (dashed lines) in terms of both direction and length. This is because the model only has the flexibility of tuning the prototype vectors, but the fixed features of the noise trigger samples cannot be effectively separated by only tuning the prototypes without deteriorating the original classification accuracy. It is also seen from Figs. \ref{visualization_2D} (e) and (f) that FTLL keeps the feature extraction of training samples unchanged. We further observe from Table \ref{quantitative} that under FTLL, $\mathcal{L}_\text{TWL} = \|\hat{W}_\text{P}-\tilde{W}_\text{P}\|_2^2$. However, this method mainly suffers from not being able to successfully embed the backdoor.

\textbf{FTAL.} It can be seen from Figs. \ref{visualization_2D} (c) and (g) that FTAL slightly alters the prototype vectors but significantly alter the feature distribution. The corresponding average PFL reaches $29061$. Therefore, from the viewpoint of deep fidelity, the performance of FTAL is unsatisfactory. Recall Fig. \ref{demo_feature}, if embedding from scratch is applied, then feature extraction and decision boundaries after convergence are determined by the specific initialization and data shuffling, while the mixed trigger samples are poisonous to the original learning process. In this situation, there is no reference model, neither the learning accuracy based fidelity nor deep fidelity can be measured.

\textbf{{FTAL\texorpdfstring{$+$}{}TWL}.} By adding the TWL to the FTAL method, we can observe from Fig. \ref{visualization_2D} (h) that the alteration of feature distribution has been well pushed towards the situation of the host model in Fig. \ref{visualization_2D} (e), while in Fig. \ref{visualization_2D} (d), the prototype vectors are still slightly different from the host model. Quantitatively, in Table \ref{quantitative}, the average PFL has been reduced from $29061$ to $687$, a significant improvement. However, this is still not in the strict sense of deep fidelity.

\textbf{FixLL.} The results of FixLL in Fig. \ref{visualization_2D} (i) is very similar to those in Fig. \ref{visualization_2D} (g) under FTAL. The reason lies in their only difference in the last layer. The quantitative results in Table \ref{quantitative} can verify this phenomenon. Since all FixLL-based methods have $\hat{W}_\text{P}=\tilde{W}_\text{P}$, their prototype vectors are identical to those of the host model and are hence not presented for brevity.

\textbf{FixLL\texorpdfstring{$+$}{}TWL.} The performance of FixLL$+$TWL is similar to that of FTAL$+$TWL except for the unaltered decision boundaries, as can be seen from Figs. \ref{visualization_2D} (j), (h) and Table \ref{quantitative}. Generally, none of the above methods can effectively satisfy deep fidelity, mainly because of the inability to preserve feature representation of the host model.

\textbf{FixLL\texorpdfstring{$+$}{}PFL.} It can be seen from Figs. \ref{visualization_2D} (k) and (e) that FixLL$+$PFL can well preserve the feature representation of training data and at the same time successfully embed all the trigger samples. From Table \ref{quantitative} we see that the average PFL is $0.4031$ while the average SPL is $0.0002$, both are significantly smaller than the former competitors. The validation accuracy is also seen to be very close to that of the host model. Therefore, the proposed FixLL$+$PFL method can simultaneously achieve successful backdoor embedding and deep fidelity.

\textbf{FixLL\texorpdfstring{$+$}{}SPL.}The performance of FixLL$+$SPL is seen from Fig. \ref{visualization_2D} (l) to be similar to that of the FixLL$+$PFL method. In Table \ref{quantitative}, The PFL of FixLL$+$SPL is slightly higher than that of FixLL$+$PFL, in that the former does not directly penalize the inconsistency of penultimate features. 

\begin{table}[!t]
	\renewcommand{\arraystretch}{1.}
	\caption{Quantitative measurements of watermarked models in Figs. \ref{visualization_2D} and \ref{visualization_3D} for MNIST classification.}
	\centering
	\setlength{\tabcolsep}{3pt}
	\label{quantitative}
	{\small{\begin{tabular}{c|l|ccrrrr}
		\hline
		\hline
		Dimension & Model & Val acc & Trig acc & \multicolumn{1}{c}{$\mathcal{L}_\text{TWL}$} & \multicolumn{1}{c}{Ave $\mathcal{L}_{\text{PFL}|\mathcal{X}}$} & Ave $\mathcal{L}_{\text{SPL}|\mathcal{X}}$ & $\|\hat{W}_\text{P} - \tilde{W}_\text{P} \|^2_2$\\ 
		\hline
		\multirow{8}{*}{\begin{tabular}{@{}c@{}}ResNet18\\MNIST\\$d=2$\end{tabular}} & Host & $99.32\%$ & $10\%$ & $-$ & $-$ & $-$ &$-$\\
		&FTLL & $99.11\%$ & $10\%$ & $3.8472$ & $0.0000$ & $0.4466$ &$3.8472$\\		
		&FTAL & $99.24\%$ & $100\%$ & $752.3854$ & $29061.8281$& $0.0746$ & $0.0165$\\		
		&FTAL+TWL & $99.29\%$ & $100\%$ & $5.9824$ & $687.8589$ & $0.0122$ & $0.0805$\\
		&FixLL & $99.16\%$ & $100\%$ & $696.4334$ & $20755.1348$ & $0.0464$ & $0.0000$\\		
		&FixLL+TWL & $99.29\%$ & $100\%$ & $6.0605$ & $873.3782$ & $0.0120$ & $0.0000$\\		
		&FixLL+PFL & $\mathbf{99.30}\%$ & $100\%$ & $337.7948$ & $\mathbf{0.4031}$ & $\mathbf{0.0002}$ & $0.0000$\\		
		&FixLL+SPL & $\mathbf{99.36}\%$ & $100\%$ & $806.9566$ & $\mathbf{11.5167}$ & $\mathbf{0.0005}$ & $0.0000$\\
		\hline
		\multirow{8}{*}{\begin{tabular}{@{}c@{}}ResNet18\\MNIST\\$d=3$\end{tabular}} & Host & $99.63\%$ & $10\%$ & $-$ & $-$ & $-$ &$-$\\
		&FTLL & $98.94\%$ & $8\%$ & $2.7318$ & $0.0000$ & $0.0953$ &$2.7318$\\
		&FTAL & $99.52\%$ & $100\%$ & $392.8843$ & $1102.6996$& $0.0057$ & $0.0507$\\
		&FTAL+TWL & $99.51\%$ & $100\%$ & $6.4869$ & $56.5022$ & $0.0027$ & $0.2520$\\
		&FixLL & $99.49\%$ & $100\%$ & $421.3672$ & $1733.9900$ & $0.0064$ & $0.0000$\\
		&FixLL+TWL & $99.48\%$ & $100\%$ & $7.1650$ & $64.0241$ & $0.0021$ & $0.0000$\\
		&FixLL+PFL & $\mathbf{99.57}\%$ & $100\%$ & $624.5971$ & $\mathbf{0.6379}$ & $\mathbf{0.0003}$ & $0.0000$\\
		&FixLL+SPL & $\mathbf{99.64}\%$ & $100\%$ & $563.0665$ & $\mathbf{13.2050}$ & $\mathbf{0.0003}$ & $0.0000$\\
		\hline
		\hline
	\end{tabular}}}
\end{table}

\begin{table}[!t]
	\renewcommand{\arraystretch}{1.}
	\caption{Quantitative measurements of watermarked models for CIFAR-10 and CIFAR-100 classification.}
	\centering
	\setlength{\tabcolsep}{2pt}
	\label{quantitative_CIFAR10}
	{\small{\begin{tabular}{c|l|ccrrrr}
		\hline
		\hline
		Dimension & Model & Val acc & Trig acc & \multicolumn{1}{c}{$\mathcal{L}_\text{TWL}$} & \multicolumn{1}{c}{Ave $\mathcal{L}_{\text{PFL}|\mathcal{X}}$} & Ave $\mathcal{L}_{\text{SPL}|\mathcal{X}}$ & $\|\hat{W}_\text{P} - \tilde{W}_\text{P} \|^2_2$\\ 
		\hline
		\multirow{8}{*}{\begin{tabular}{@{}c@{}}ResNet18\\CIFAR-10\\$d=64$\end{tabular}} & Host & $86.44\%$ & $11\%$ & $-$ & $-$ & $-$ &$-$\\
		&FTLL & $84.98\%$ & $16\%$ & $25.9460$ & $0.0000$ & $0.0519$ &$25.9560$\\		
		&FTAL & $86.36\%$ & $100\%$ & $378.5653$ & $6.2938$& $0.0373$ & $0.4482$\\		
		&FTAL+TWL & $86.04\%$ & $100\%$ & $7.4755$ & $2.8301$ & $0.0306$ & $0.2899$\\
		&FixLL & $\mathbf{86.48}\%$ & $100\%$ & $400.0898$ & $17.2949$ & $0.0433$ & $0.0000$\\		
		&FixLL+TWL & $86.10\%$ & $100\%$ & $7.6574$ & $4.1752$ & $0.0280$ & $0.0000$\\		
		&FixLL+PFL & $\mathbf{86.45}\%$ & $100\%$ & $429.1258$ & $\mathbf{0.8709}$ & $\mathbf{0.0095}$ & $0.0000$\\		
		&FixLL+SPL & $\mathbf{86.47}\%$ & $100\%$ & $242.3010$ & $\mathbf{0.1239}$ & $\mathbf{0.0006}$ & $0.0000$\\
		\hline
		\multirow{8}{*}{\begin{tabular}{@{}c@{}}WRN28\_10\\CIFAR-100\\$d=640$\end{tabular}} & Host & $75.50\%$ & $0\%$ & $-$ & $-$ & $-$ &$-$\\
		&FTLL & $74.64\%$ & $4\%$ & $1539.3673$ & $0.0000$ & $0.0005$ &$1539.3673$\\		
		&FTAL & $74.64\%$ & $100\%$ & $25592.1826$ & $59.0712$& $0.0020$ & $64.2557$\\		
		&FTAL+TWL & $72.23\%$ & $100\%$ & $275.0834$ & $33.2550$ & $0.0050$ & $21.1531$\\
		&FixLL & $75.36\%$ & $100\%$ & $18419.1835$ & $98.0179$ & $0.0017$ & $0.0000$\\		
		&FixLL+TWL & $72.41\%$ & $100\%$ & $280.7973$ & $28.2255$ & $0.0057$ & $0.0000$\\		
		&FixLL+PFL & $\mathbf{75.73}\%$ & $100\%$ & $25915.0568$ & $\mathbf{1.5136}$ & $\mathbf{0.0002}$ & $0.0000$\\		
		&FixLL+SPL & $\mathbf{75.77}\%$ & $100\%$ & $10864.1162$ & $\mathbf{16.3229}$ & $\mathbf{0.0009}$ & $0.0000$\\
		\hline
		\hline
	\end{tabular}}}
\end{table}

\textbf{Additional Remarks.} \textbf{i)} In Table \ref{quantitative}, the column containing validation accuracies indicates that backdoor embedding will inevitably reduce the validation accuracy of the model in most situations. However, the only exception in both the $2$- and $3$-dimensional cases is the FixLL$+$SPL method, whose validation accuracies improved slightly over the host model. We hypothesize that the reason lies in the direct control of output probability via the SPL. Besides, The validation accuracies of FixLL$+$PFL are the second-highest among the competitors. \textbf{ii)} The TWL has been shown in Figs. \ref{visualization_2D} and \ref{visualization_3D} to be effective in regularizing the feature distributions and pushing them towards the host model compared to the methods without using it. However, as a multimedia-watermarking-analogous measurement, a high TWL value does not necessarily mean low fidelity or low deep fidelity, because it can be seen from Table \ref{quantitative} when $d=2$, FTAL can have a validation accuracy of $99.24\%$ when $\mathcal{L}_\text{TWL}$ is $752$, but FTAL$+$TWL suffers from a large feature loss of $688$ although $\mathcal{L}_\text{TWL}$ is only $5.9824$. Similar phenomena can be seen in other methods and when $d=3$. \textbf{iii)} The $3$-dimensional version ($d=3$) of ResNet18 turns out to be more effective on the MNIST dataset than the $2$-dimensional version for its slightly increased validation accuracies across all the implemented methods.

\subsection{CIFAR-10 and CIFAR-100 Results}
Additional quantitative results of the watermarked models for CIFAR-10 (ResNet18 in Table \ref{model}) and CIFAR-100 (WRN28\_10) classification tasks are summarized in Table \ref{quantitative_CIFAR10}. In terms of backdoor embedding, performance results are consistent with those presented in Table \ref{quantitative}. Additionally, it can be seen from the Tables \ref{quantitative} and \ref{quantitative_CIFAR10} that for the methods in Group $\textbf{B}$ and under a fixed number of classes of $10$, increasing the feature dimension effectively results in the reduction of the PFL, indicating that the feature clusters tend to move to the origin in a higher-dimensional space. However, by using the PFL and SPL, i.e., the proposed methods in Group \textbf{C}, the PFL becomes insensitive to the feature dimension. For the more challenging task of CIFAR-100 classification, the penultimate feature dimension is $d=640$, resulting in larger TWL values, but via the use of the proposed embedding methods, both PFL and SPL can be reduced to very small value ranges, indicating preserved feature learning on the training set, with unchanged decision boundaries. 

\subsection{Performance Analysis and Hyperparameter Tuning}
\subsubsection{Convergence} 
\begin{figure}[!t]
	\centering
	\subfigure[Val. and Trig. acc.]{\includegraphics[width=2.5in]{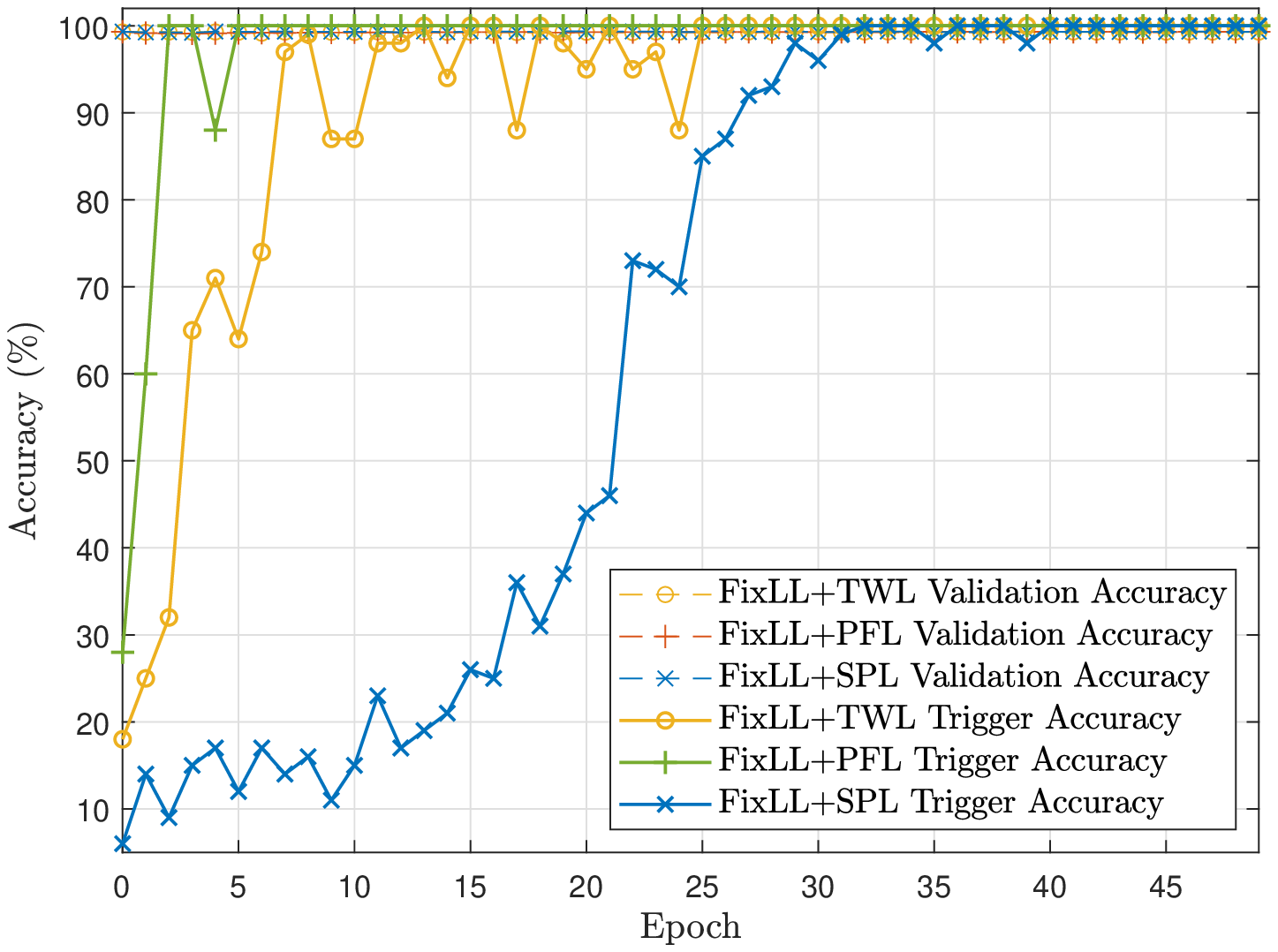}}
	\subfigure[TWL.]{\includegraphics[width=2.5in]{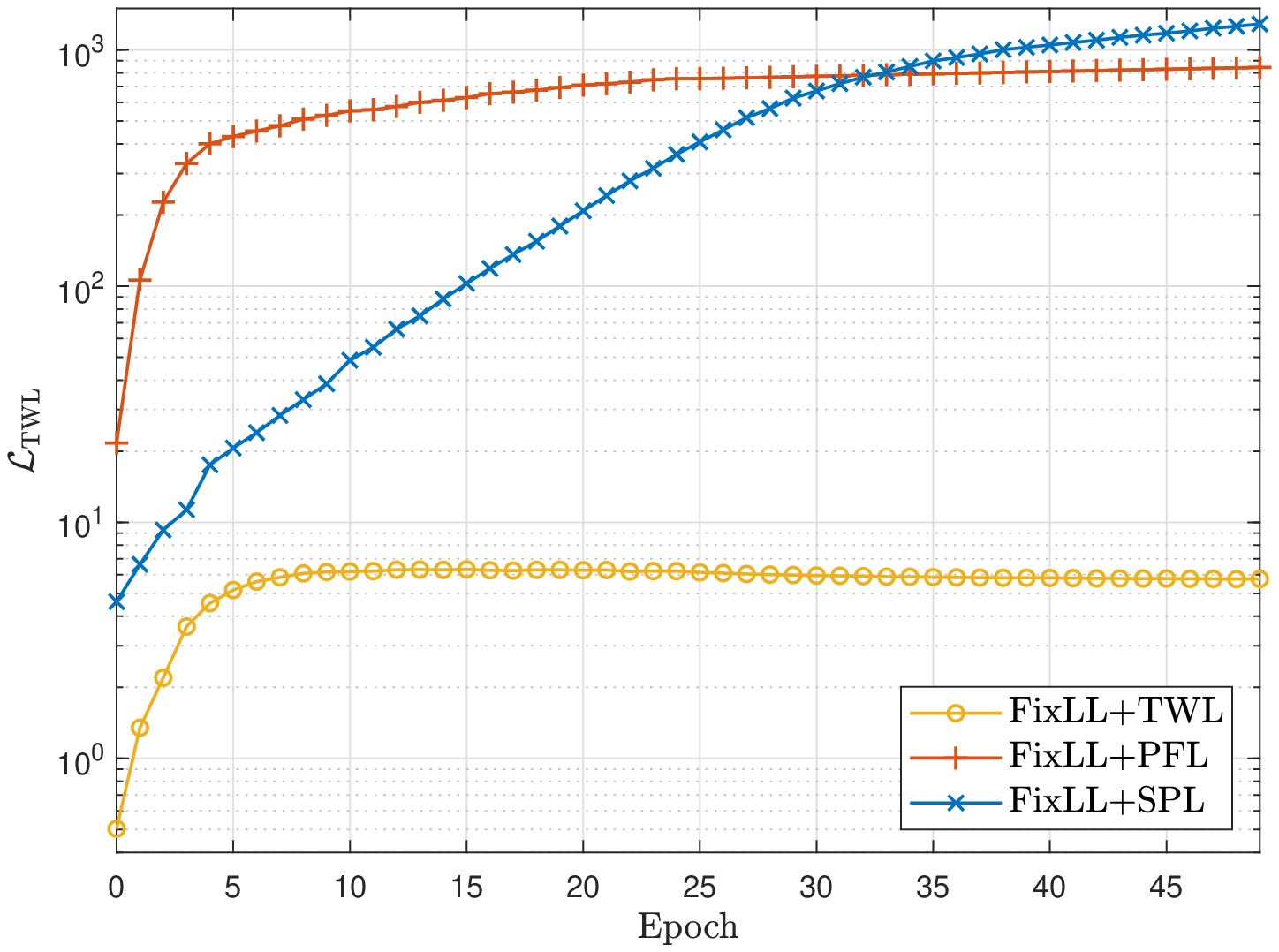}}
	\hfill\\
	\subfigure[PFL.]{\includegraphics[width=2.5in]{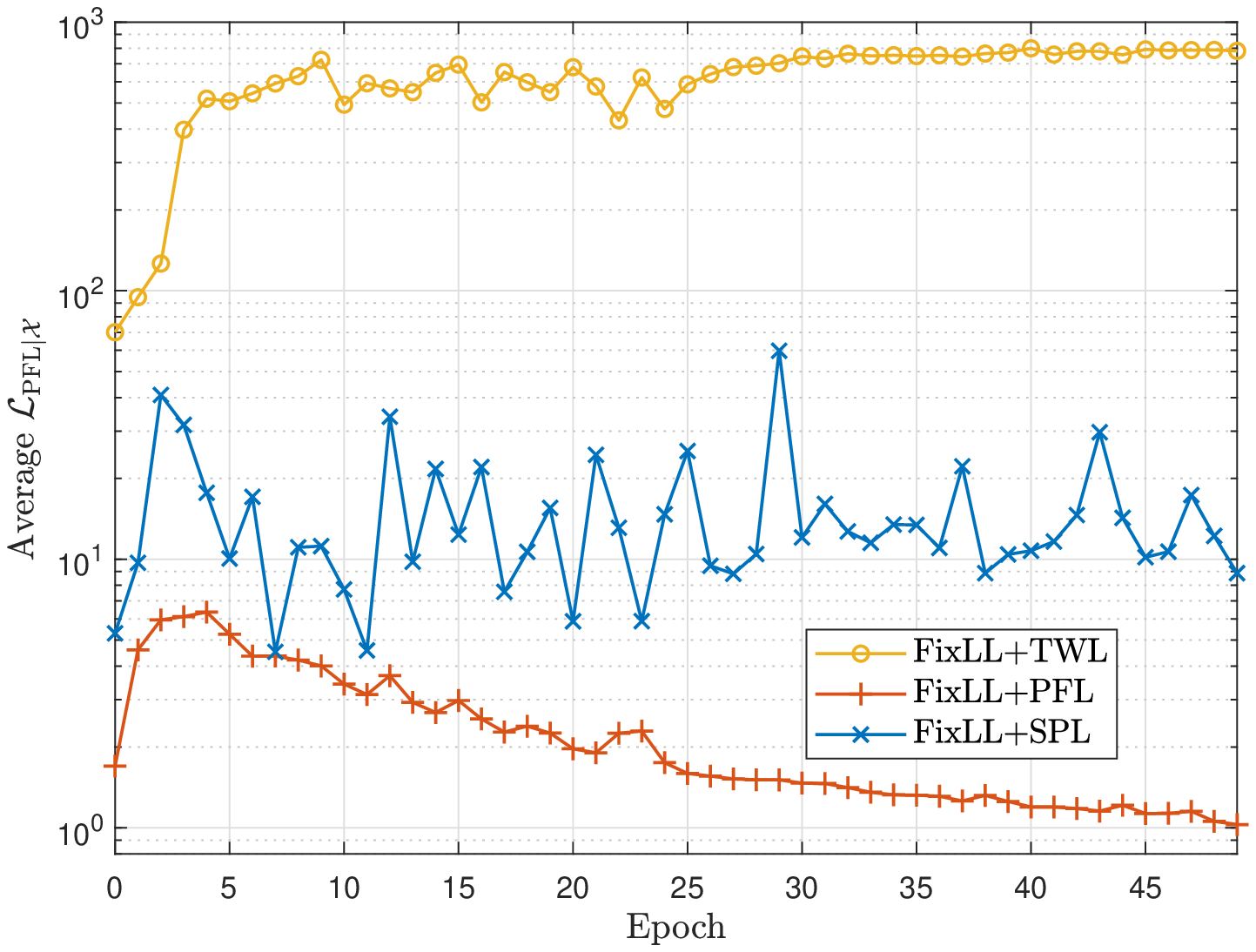}}
	\subfigure[SPL.]{\includegraphics[width=2.5in]{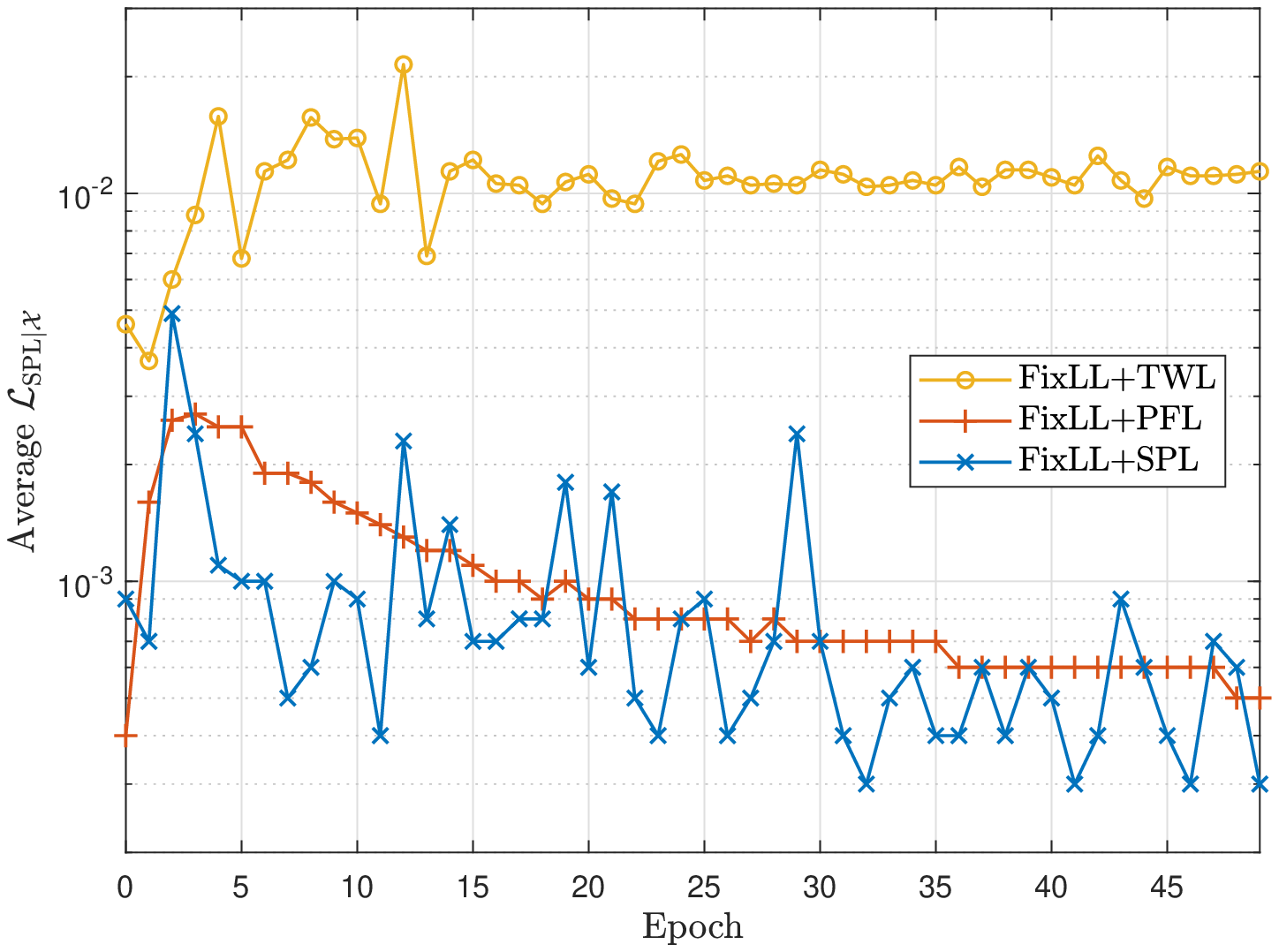}}
	\hfill\\
	\caption{Performance analysis of FixLL$+$TWL, FixLL$+$PFL, and FixLL$+$SPL methods on MNIST in terms of accuracies, TWL, PFL, and SPL, as functions of epoch, where $m=4$, $\alpha=\beta=0.01$, $\gamma=1000$.}
	\label{loss_curves}
\end{figure}

Since fine-tuning-based backdoor watermark embedding is initialized by setting $\tilde{W} = \hat{W}$, the losses, including $\mathcal{L}_\text{TWL}$, $\mathcal{L}_{\text{PFL}|\mathcal{X}}$, $\mathcal{L}_{\text{SPL}|\mathcal{X}}$, and $\|\hat{W}_\text{P} - \tilde{W}_\text{P} \|^2_2$, actually are all zero during initialization, but the trigger set is not learned at this stage. The embedding process therefore establishes a trade-off among trigger accuracy, learning accuracy, and these losses in the respective methods. Fig. \ref{loss_curves} shows how these quantities change over epochs, where the models in FixLL$+$TWL and FixLL$+$PFL are fine-tuned for $20$ epochs, but the model in FixLL$+$SPL is fine-tuned for $50$ epochs. The reason is demonstrated in Fig. \ref{loss_curves} (a), in which both the validation and trigger accuracies of the FixLL$+$TWL and FixLL$+$PFL methods approach $100\%$ after $10$ epochs, but this happens after $30$ epochs for the FixLL$+$SPL method. Since FixLL$+$SPL and FixLL$+$PFL do not have active control of the weight loss, they have much higher TWL values than FixLL$+$TWL, which can be seen in Fig. \ref{loss_curves} (b). From Figs. \ref{loss_curves} (c) and (d), we observe that FixLL$+$PFL is the best method in preserving feature distribution in terms of both measurements, followed by FixLL$+$SPL and then the FixLL$+$TWL method. The value ranges of the y-axes in Figs. \ref{loss_curves} (b)\---(d) indicate how the values of $\alpha$ in (\ref{FTAL+TWL_Embedding}) and (\ref{FixLL+TWL_Embedding}), $\beta$ in (\ref{FixLL+PFL}), and $\gamma$ in (\ref{FixLL+SPL}) are selected. 

\begin{figure}[!t]
	\centering
	\subfigure[FixLL$+$TWL Accuracies.]{\includegraphics[width=2.5in]{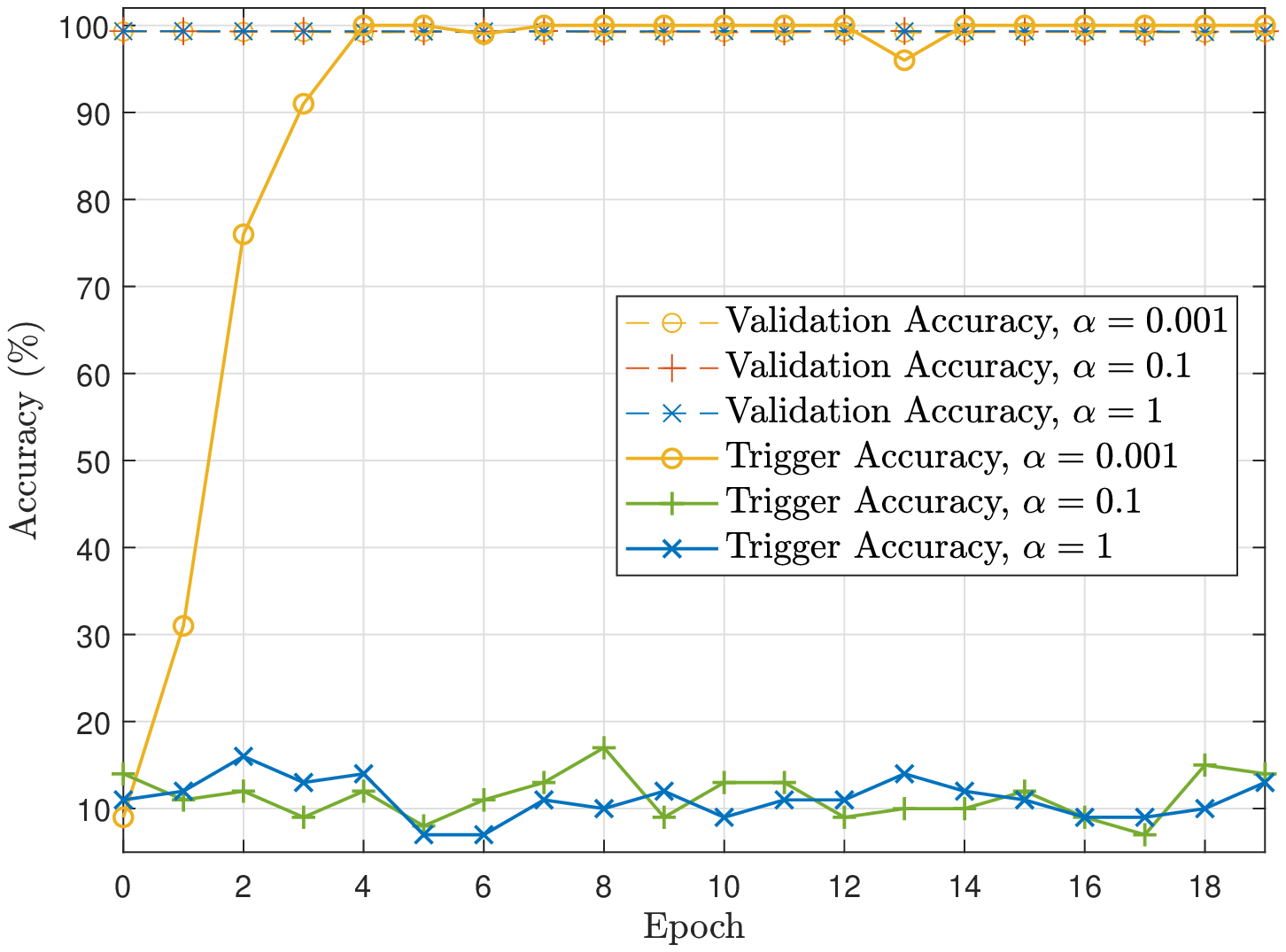}}
	\subfigure[FixLL$+$TWL PFL.]{\includegraphics[width=2.5in]{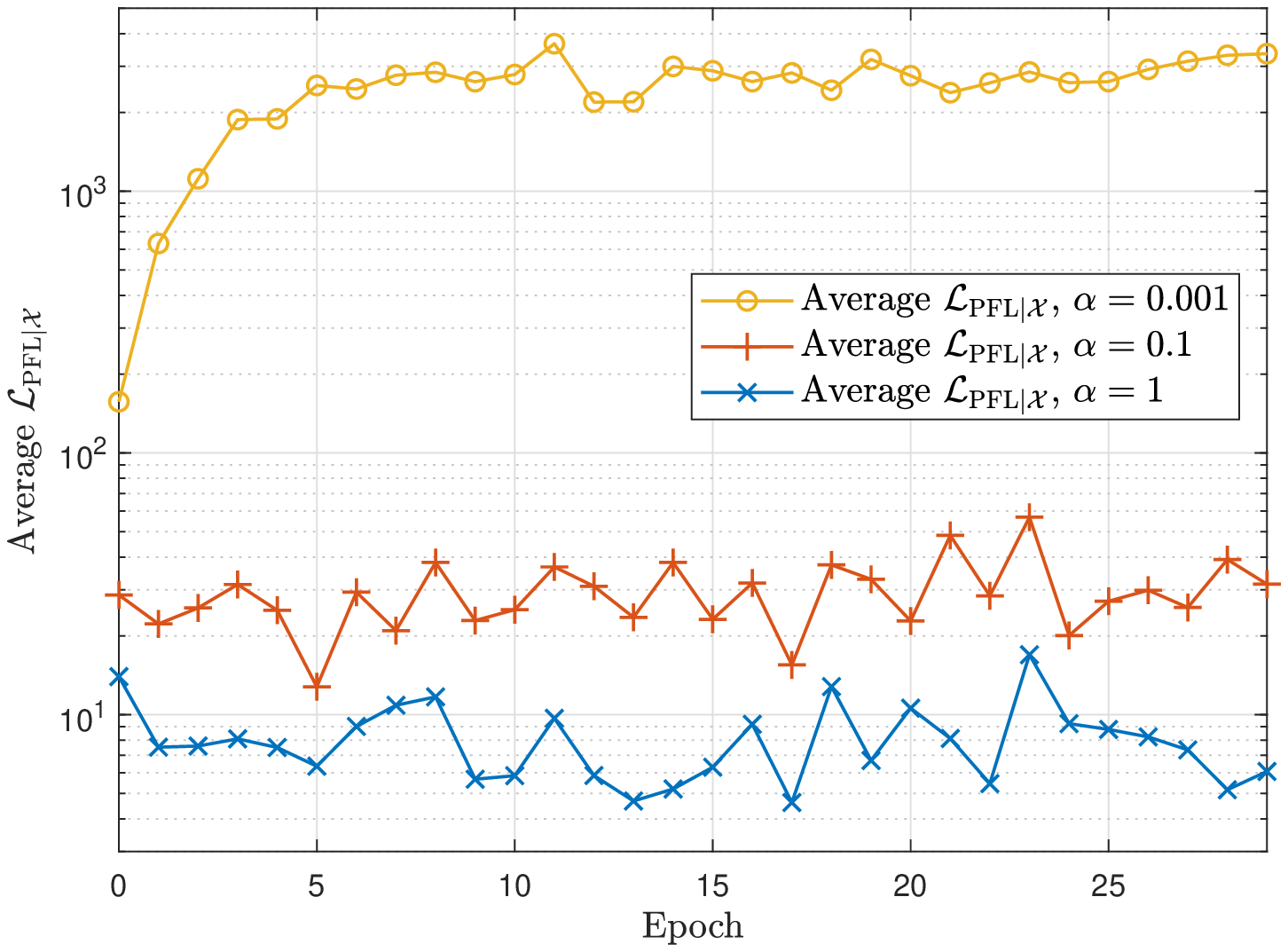}}
	\hfill\\
	\subfigure[FixLL$+$PFL Accuracies.]{\includegraphics[width=2.5in]{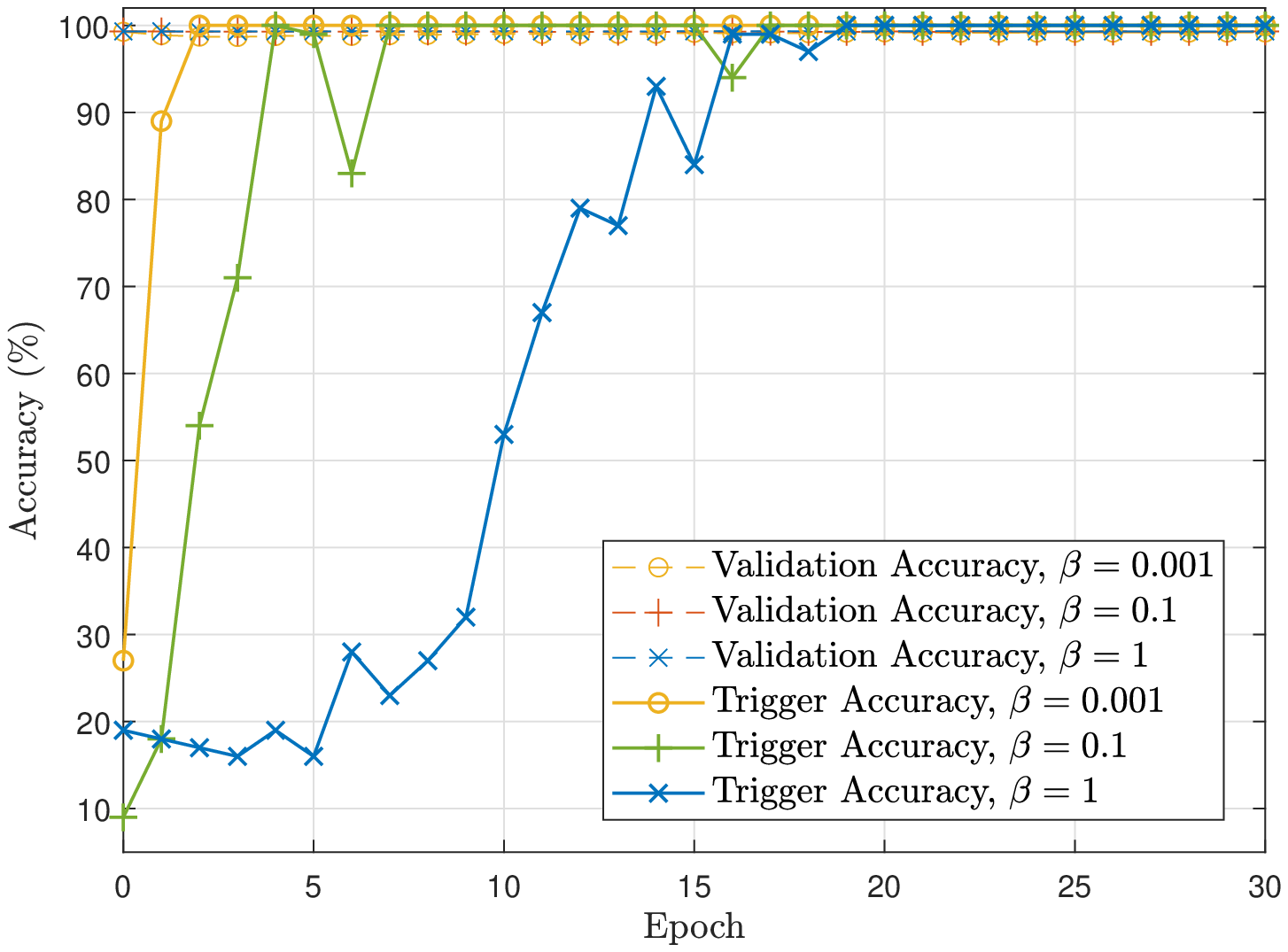}}
	\subfigure[FixLL$+$PFL PFL.]{\includegraphics[width=2.5in]{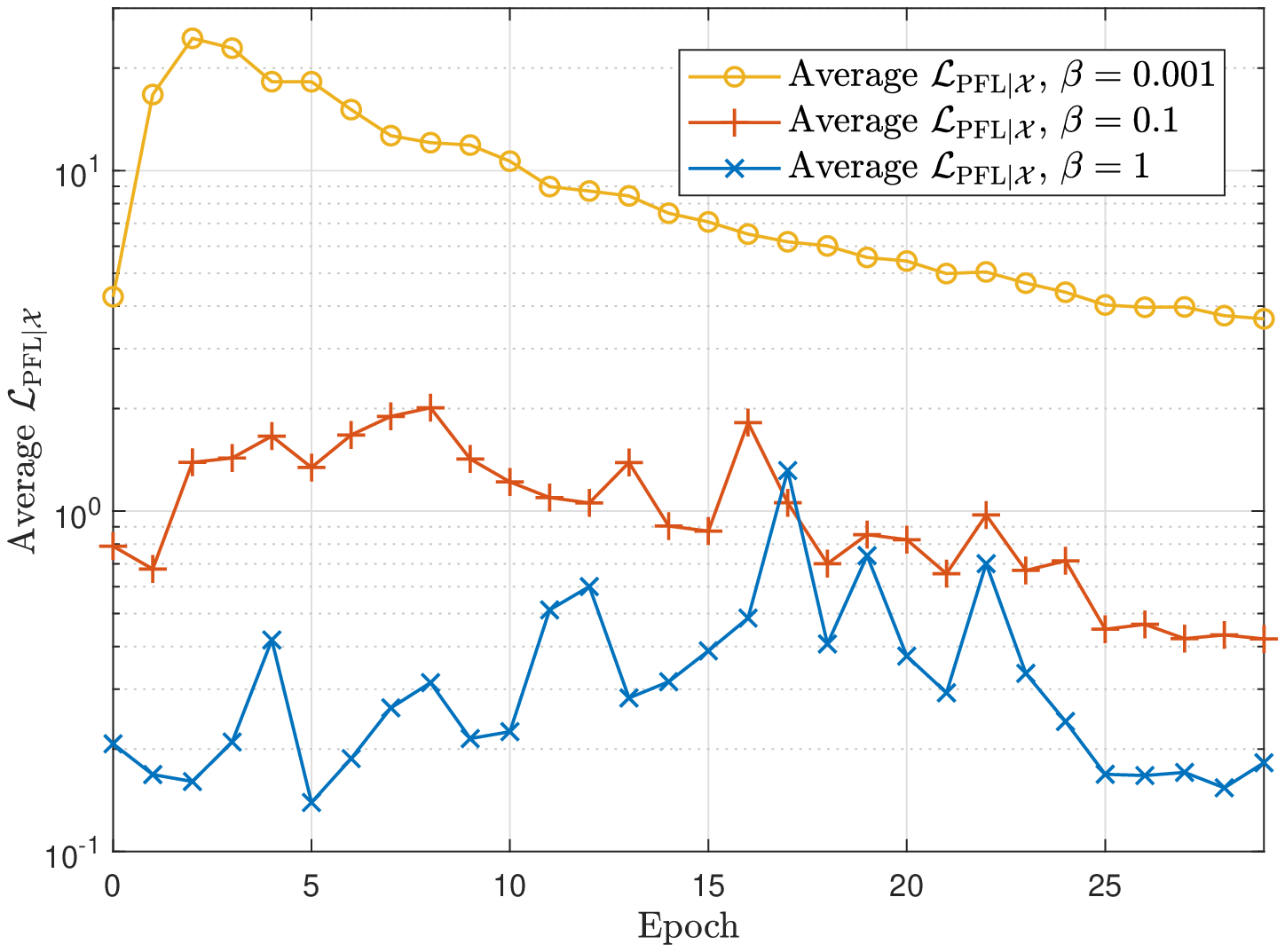}}
	\hfill\\
	\subfigure[FixLL$+$SPL Accuracies.]{\includegraphics[width=2.5in]{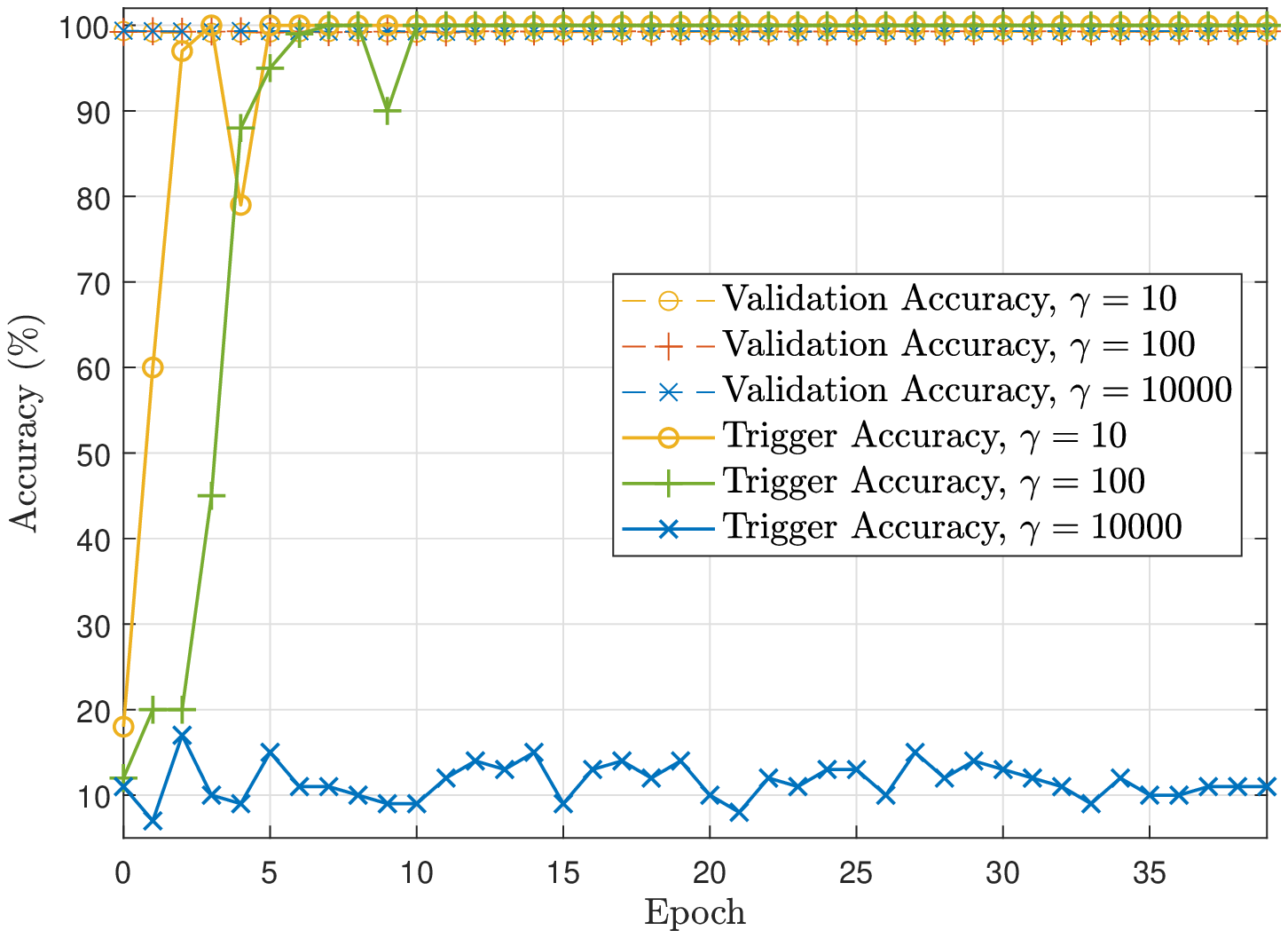}}
	\subfigure[FixLL$+$SPL PFL.]{\includegraphics[width=2.5in]{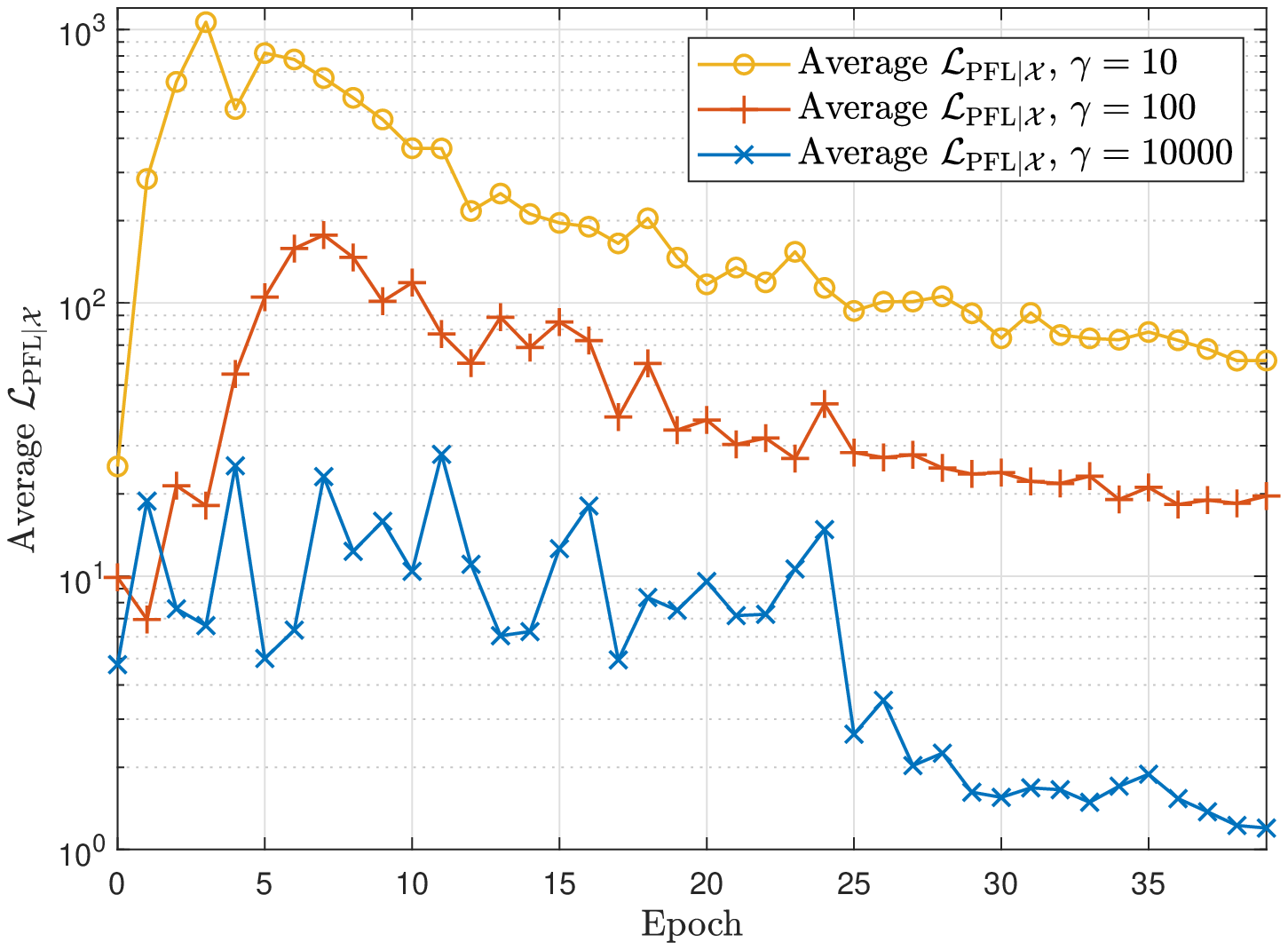}}
	\hfill\\
	\caption{Performance analysis of FixLL$+$TWL, FixLL$+$PFL, and FixLL$+$SPL methods on MNIST under different values of $\alpha$, $\beta$, and $\gamma$, where $m=4$, measured by validation accuracy, trigger accuracy, and PFL.}
	\label{hyper_tune}
\end{figure}

\begin{table}[!t]
	\renewcommand{\arraystretch}{1.}
	\caption{Quantitative performance measurements under different values of $m$ for MNIST classification, where batch size is $32$, $d=2$, $\alpha=\beta=0.01$, and $\gamma=1000$.}
	\centering
	\setlength{\tabcolsep}{6pt}
	\label{m_per_batch}
	\begin{tabular}{c|l|ccr}
		\hline
		\hline
		$m$  & Model & Validation acc & Trigger acc  & Average $\mathcal{L}_{\text{PFL}|\mathcal{X}}$\\ 
		\hline
		$0$ & Host & $99.32\%$ & $10\%$ & $-$\\	
		\hline	
		\multirow{3}{*}{$4$} & FixLL+TWL & $99.29\%$ & $100\%$ & $873.3782$\\		
		&FixLL+PFL & ${99.30}\%$ & $100\%$ & ${0.4031}$\\		
		&FixLL+SPL & ${99.36}\%$ & $100\%$ & ${11.5167}$\\
		\hline	
		\multirow{3}{*}{$16$} & FixLL+TWL & $99.31\%$ & $100\%$ & $965.2145$\\		
		&FixLL+PFL & ${99.31}\%$ & $100\%$ & ${1.7237}$\\		
		&FixLL+SPL & ${99.32}\%$ & $100\%$ & ${17.2922}$\\
		\hline	
		\multirow{3}{*}{$32$} & FixLL+TWL & $99.27\%$ & $100\%$ & $1202.4749$\\		
		&FixLL+PFL & ${99.29}\%$ & $100\%$ & ${1.4907}$\\		
		&FixLL+SPL & ${99.34}\%$ & $100\%$ & ${15.0654}$\\
		\hline
		\hline
	\end{tabular}
\end{table} 

\subsubsection{\texorpdfstring{$\alpha$}{}, \texorpdfstring{$\beta$}{}, and \texorpdfstring{$\gamma$}{}}
Performance measurements under different weighting hyperparameters $\alpha$, $\beta$, and $\gamma$, are summarized in Fig. \ref{hyper_tune}, where the left column of shows validation and trigger accuracies after backdoor embedding and the right column shows the PFL which directly measures the alteration of training data feature representation. It can be seen from Fig. \ref{hyper_tune} (a) and (\ref{FixLL+TWL_Embedding}) that $\alpha$ should be set relatively small to emphasize the learning of trigger samples. Otherwise, e.g., when $\alpha=0.1$ or $1$, the trigger samples cannot be learned, leading to the failure of backdoor embedding. Similarly, increasing $\beta$ and $\gamma$ in their respective scales will slow down or also disable the learning of trigger samples, as can be seen from Figs. \ref{hyper_tune} (c) and (e). However, according to Figs. \ref{hyper_tune} (b), (d), and (f), increasing these weighting hyperparameters can effectively reduce the alteration of feature representation. Therefore, these hyperparameters control the trade-off between backdoor learning and deep fidelity. Among the three methods, FixLL$+$PFL is seen to be the most effective in preserving feature representation, thanks to the direct feature control via the PFL. 

\subsubsection{\texorpdfstring{$m$}{} per Batch}
We now examine how the mix hyperparameter $m$ affects the performance of the watermarking schemes, and the results are presented in Table \ref{m_per_batch}. Since the number of trigger samples is much less than the number of training samples, backdoor embedding can be considered as a learning process with data augmentation, i.e., the trigger samples are over-sampled such that the ratio between the trigger samples (repeatedly seen per epoch) and training samples (seen once per epoch) are $m/32$ under the fixed training data batch size of $32$, and the actual batch size becomes $32+m$. For example, consider the MNIST dataset with $60000$ training samples and $100$ trigger samples for backdoor learning, then $m=4$ indicates that the number of trigger samples is over-sampled from $100$ to $7500$ via random repetition in the embedding process. From Table \ref{m_per_batch}, it can be seen that the proposed FixLL$+$PFL and FixLL$+$SPL methods are insensitive to $m$, because of the disentanglement of backdoor learning and feature preservation in both schemes. However, since the FixLL$+$TWL method considers the mixed training and trigger set as a whole dataset, increasing $m$ lets the model learns more of the trigger samples and resulted in the increase of the PFL.

\begin{figure}[!t]
	\centering
	\subfigure[Noise.]{\includegraphics[width=1.5in,height=1.5in]{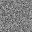}}
	\subfigure[Abstract.]{\includegraphics[width=1.5in, height=1.5in]{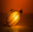}}
	\subfigure[Abstract+String.]{\includegraphics[width=1.5in,height=1.5in]{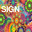}}
	\hfill\\
	\caption{Examples of $32 \times 32$ trigger samples. (b) and (c) are based on the abstract images used in \cite{Backdoor_2018_turning_weakness}.}
	\label{Instance}
\end{figure}

\subsubsection{Trigger Diversity}
Considering the diversity of trigger samples and in addition the noise samples used in previous experiments, we present additional experimental results using both abstract images \cite{Backdoor_2018_turning_weakness} and string based images as the trigger set. Examples of such triggers are presented in Fig. \ref{Instance}, and experimental results are summarized in Table \ref{Instance_Results}. It can be seen that the results are consistent with those in Table \ref{quantitative_CIFAR10} as FixLL+PFL and FixLL+SPL yield the best performance in terms of learning accuracy and deep fidelity, thus the performance is insensitive to the choice of trigger samples.

\begin{table}[!t]
	\renewcommand{\arraystretch}{1.1}
	\setlength{\tabcolsep}{6pt}
	\caption[]{Quantitative results of ResNet18 ($d=64$) models watermarked by ``Abstract'' and ``Abstract+String'' trigger samples for CIFAR-10 classification.}
	\label{Instance_Results}
	\centering
	\begin{tabular}{c|l|ccrr}
		\hline
		\hline
		& \multicolumn{1}{c|}{{Model}} & Val acc & Trig acc & Ave $\mathcal{L}_{\text{PFL}|\mathcal{X}}$ & $\|\hat{W}_\text{P} - \tilde{W}_\text{P} \|^2_2$\\
		\hline
		& Host & $86.44\%$ & $11\%$ & $-$ & $-$ \\
		\hline
		\parbox[t]{3mm}{\multirow{6}{*}{\rotatebox[origin=c]{90}{Abstract}}} 
		& FTAL & $86.11\%$ & $100\%$ & $6.3216$ & $0.3696$\\
		& FTAL+TWL & $86.12\%$  & $100\%$  & $2.9624$ & $0.5584$\\
		& FixLL & $86.34\%$  & $100\%$  & $16.2069$ & $0.0000$\\
		& FixLL+TWL & $86.29\%$  & $100\%$  & $6.4524$ & $0.0000$\\
		& FixLL+PFL & $86.21\%$  & $100\%$  & $0.4170$ & $0.0000$\\
		& FixLL+SPL  & $86.42\%$  &  $100\%$  & $0.4261$ & $0.0000$\\
		\hline
		\parbox[t]{3mm}{\multirow{6}{*}{\rotatebox[origin=c]{90}{Abstract+String}}} 
		& FTAL & $85.99\%$ & $100\%$ & $10.1786$ & $0.5028$\\
		& FTAL+TWL & $86.23\%$ & $100\%$  & $2.9295$ & $0.4194$\\
		& FixLL & $86.25\%$  & $100\%$  & $15.7308$ & $0.0000$\\
		& FixLL+TWL & $86.37\%$  & $100\%$  & $6.0397$ &  $0.0000$\\
		& FixLL+PFL & $86.43\%$  & $100\%$  & $0.2749$ &  $0.0000$\\
		& FixLL+SPL  & $86.41\%$  & $100\%$  & $0.4723$ & $0.0000$\\
		\hline
		\hline
	\end{tabular}
\end{table}

\begin{table}[!t]
	\renewcommand{\arraystretch}{1.2}
	\centering
	\caption{Results on robustness against fine-tuning and compression using WRN28\_10 on CIFAR-100.}
	\setlength{\tabcolsep}{4pt}
	\label{Robustness}
	\begin{tabular}{c|c|ccrr}
		\hline
		\hline
		Attack & \begin{tabular}{@{}c@{}}Embedding\\Method\end{tabular} & Val acc & Trig acc & Ave $\mathcal{L}_{\text{PFL}|\mathcal{X}}$ & $||{\hat W_{\text{P}}} - {\tilde W_{\text{P}}}||_2^2$\\
		\hline
		\multirow{3}{*}{\begin{tabular}{@{}c@{}}Fine-Tune\\After $50$ epochs\end{tabular}} & Scratch & $74.54\%$ & $100\%$ & $-$ & $-$ \\
		& FixLL+Fine-Tune & $75.24\%$ & $1.2\%$ & $72.1606$ & $0.0000$\\
		& FixLL+Scratch & $75.42\%$  & $100\%$ & $102.3892$ & $0.0000$\\
		\hline
		\multirow{4}{*}{\begin{tabular}{@{}c@{}}Truncation\\$32$- to $16$-bit\end{tabular}} & Scratch & $74.67\%$ & $100\%$ & $-$ & $-$ \\
		& FixLL+Fine-Tune & $74.77\%$ & $100\%$ & $57.2526$ & $0.0000$ \\
		& FixLL+PFL & $75.56\%$ & $100\%$ & $5.2671$ & $0.0000$ \\
		& FixLL+SPL & $75.63\%$ & $100\%$ & $10.3229$ & $0.0000$ \\
		\hline
		\multirow{4}{*}{\begin{tabular}{@{}c@{}}Pruning\\$20\%$\end{tabular}} & Scratch & $74.30\%$ & $100\%$ & $-$ & $-$ \\
		& FixLL+Fine-Tune & $74.50\%$ & $100\%$ & $57.0876$ & $0.0000$ \\
		& FixLL+PFL & $75.39\%$ & $100\%$ & $6.0037$ & $0.0000$ \\
		& FixLL+SPL & $75.49\%$ & $100\%$ & $10.8788$ & $0.0000$ \\
		\hline
		\multirow{4}{*}{\begin{tabular}{@{}c@{}}Pruning\\$50\%$\end{tabular}} & Scratch & $54.76\%$ & $17.50\%$ & $-$ & $-$ \\
		& FixLL+Fine-Tune & $57.61\%$ & $23.30\%$ & $84.3395$ & $0.0000$ \\
		& FixLL+PFL & $55.04\%$ & $10.70\%$ & $43.7739$ & $0.0000$ \\
		& FixLL+SPL & $52.72\%$ & $13.40\%$ & $49.3189$ & $0.0000$ \\
		\hline
		\hline
	\end{tabular}
\end{table}

\subsubsection{Robustness against Fine-tuning and Compressiopn}
We further present additional experimental results on the robustness of the proposed methods against white-box fine-tuning and compression attacks, using both embedding from scratch and fine-tuning based methods, summarized in Table \ref{Robustness}. One result confirmed with that in \cite{Backdoor_2018_turning_weakness} is that embedding from scratch is more robust than fine-tuning based embedding to fine-tuning attacks \--- the former still yields $100\%$ trigger accuracy after $50$ epochs of fine-tuning using normal training samples, while the trigger accuracy of the latter drops to $1.2\%$. However, it is also discovered that the proposed FixLL can effectively work with embedding from scratch to gain robustness against fine-tuning attacks and at the same time preserve deep fidelity in terms of PFL and decision boundaries. Under model compression attacks, the proposed methods have exhibited robustness for reasonable compressions, while under $50\%$ heavy pruning, both Val acc and Trig acc have dropped substantially. We note that the main focus of this paper lies in fidelity, while it is possible to investigate the joint improvement of deep fidelity and robustness in future. 

In summary, the proposed embedding schemes offer a set of schemes for designers to trade-off fidelity and robustness while ensuring full backdoor learning. If the model is distributed in black-box (white-box attack is unavailable), then fine-tuning based embedding schemes incorporating FixLL and PFL (or SPL) can achieve deep fidelity. The proposed fine-tuning based schemes can also work under transfer learning scenarios with pretrained backbone modules. If the robustness against fine-tuning is an important concern, then FixLL can be used with embedding from scratch to achieve this goal and at the same time preserve the feature learning of the training data via the supervision of the predefined decision boundaries. 

\subsubsection{Limitations}
The limitations of our proposal are summarized as follows. \textbf{(1)} Deep fidelity has to be achieved via fine-tuning-based embedding, which has been verified to be more vulnerable than embedding from scratch to fine-tuning attacks, as reflected in the results shown in Table 8. However, as state-of-the-art deep learning models become bigger and bigger, fine-tuning-based embedding is more practical in terms of efficiency. \textbf{(2)} The proposed techniques (FixLL, PFL, and SPL) only apply to deep classification models with a backbone feature extractor and a linear classifier, because representation learning and classification are the two unique functionalities in these models. However, the extension of deep fidelity is not straightforward. First, the watermarks for other types of models (e.g., language models, graph neural networks, encoder-decoder structures, etc.) need to be designed differently, which is out of the scope of this paper. Second, other types of models have their respective modules to deal with when considering deep fidelity. For example, we need to preserve the word embedding mechanism of language models and statistical consistency of latent features for variational autoencoders, while the output layer of generative models is not a classifier. All these differences are worth of further investigation in future.

\section{Conclusion}\label{sec_6}
In this paper, we have proposed the concept of \emph{deep fidelity} for DNN backdoor watermarking, which enriches the current learning-accuracy-based fidelity definition by the additional consideration of the internal mechanisms of the model. Focusing on deep image classification models as our first attempt, we have proposed the notion that to achieve deep fidelity, we need to preserve the feature representation of training data and the decision boundaries of these models. Based on a comprehensive review of the existing backdoor watermark embedding methods, we showed that training from scratch or retraining some layers cannot ensure deep fidelity. Then, based on the existing FTAL method, we proposed several mechanisms to achieve deep fidelity. Specifically, to preserve the decision boundaries, we incorporated the result reported in \cite{Fix_Pro_2018_ICLR,Fix_Pro_2021_TNNLS} that the last layer classifier can be fixed without deteriorating classification accuracy, based on which the FixLL method is proposed. To preserve the host model functionality of feature extraction, we further proposed two loss functions, i.e., PFL in (\ref{PFL}) and SPL in (\ref{SPL}), respectively. In comparison with the existing methods as well as TWL-based multimedia-watermarking-analogous methods, the proposed methods are extensively evaluated via experiments using modified ResNet18 on MNIST and CIFAR-10, and WRN28\_10 on CIFAR-100 classification tasks.

We note that in addition to the noise, abstract, and abstract+string images considered in this paper, it is worth to comprehensively study the choice of trigger samples and the influence on deep fidelity or other criteria. These triggers can be selected between close- and open-set, natural and computer generated images, and spatial and transform domain patterns. Meanwhile, besides the backdoor embedding considered in this paper, deep fidelity can also be considered in internal embedding schemes. In these scenarios, it is critical to identify the key functionalities of the to-be-protected model in addition to learning accuracy, which can provide clues for the design of embedding mechanisms and loss functions to achieve deep fidelity. Lastly, although this paper has focused on the watermarking of deep discriminative models for image classification tasks, the proposed notion of deep fidelity can be applied to general DNN structures such as discriminative models for regression tasks and also generative models. However, the extension of deep fidelity is not straightforward. For other types of models, e.g., generative models, there is a need of dedicated watermark designs, loss functions, or other treatments to ensure the preservation of their internal mechanisms.

\bibliographystyle{elsarticle-num} 
\bibliography{nn_watermark.bib}

\end{document}